\documentclass[prx,a4paper,twocolumn,showpacs,floatfix,superscriptaddress,amsmath,amsfonts,amssymb,preprintnumbers,citeautoscript]{revtex4-2}
\pdfoutput=1
\usepackage[T1]{fontenc}\usepackage[latin1]{inputenc}
\usepackage{dcolumn,graphicx,color,booktabs,microtype,afterpage,bm} \graphicspath{{Figures/}}
\usepackage[charter,greekuppercase=italicized]{mathdesign}
\usepackage{subfigure,sidecap}
\renewcommand{\figurename}{Fig.}
\renewcommand{\tablename}{Table}

\usepackage[shortlabels]{enumitem}
\makeatletter
\newcommand{\bibstyle@supplement}{\bibpunct[, ]{[S}{]}{;}{n}{,}{, S\hspace{-2.1pt}}%
\gdef\bibnumfmt##1{[S##1]}}
\makeatother

\newcommand{\boldsigma}{{\bm \sigma}}
\newcommand{\boldtau}{{\bm \tau}}
\newcommand{\boldeta}{{\bm \eta}}
\newcommand{\boldepsilon}{{\bm \epsilon}}
\newcommand{\boldLambda}{{\bm \Lambda}}
\newcommand{\boldchi}{{\bm \chi}}

\newcommand{\be}{\begin{equation}}
\newcommand{\ee}{\end{equation}}

\newcommand{\bea}{\begin{eqnarray}}
\newcommand{\eea}{\end{eqnarray}}
\newcommand{\bd}{\begin{displaymath}}
\newcommand{\ed}{\end{displaymath}}
\newcommand{\ba}{\begin{array}}
\newcommand{\ea}{\end{array}}
\newcommand{\bi}{\begin{itemize}}
\newcommand{\ei}{\end{itemize}}
\newcommand{\bc}{\begin{center}}
\newcommand{\ec}{\end{center}}
\newcommand{\bfl}{\begin{flushleft}}
\newcommand{\efl}{\end{flushleft}}
\newcommand{\bfr}{\begin{flushright}}
\newcommand{\efr}{\end{flushright}}

\def\Ce{\rm{CeB$_6$}}
\def\U{\rm{URu$_2$Si$_2$}}

\def\ua{\uparrow}
\def\da{\downarrow}
\def\ket#1{\left\vert #1 \right\rangle}

 \def\bK{{\bf K}}
\def\bq{{\bf q}} \def\bQ{{\bf Q}}
\def\bJ{{\bf J}}
\def\bM{{\bf M}}
\def\bD{{\bf D}}
\def\bun{{\bf 1}}

\def\da{\downarrow} \def\ua{\uparrow}
 
\def\6{\partial}

\def\ss{} 
 
\def\o{\omega}  
 \def\L{\Lambda}

\def\bra{\langle}
\def\ket{\rangle}

\usepackage[colorlinks,linkcolor=blue,urlcolor=blue,citecolor=blue]{hyperref}

\begin{document}
\date{\today}

\title{Field-angle resolved magnetic excitations as a probe of hidden-order symmetry in CeB$_6$}

\author{P.~Y.~Portnichenko}
\affiliation{Institut f\"ur Festk\"orper- und Materialphysik, Technische Universit\"at Dresden, 01069 Dresden, Germany}

\author{A.~Akbari}\email[Corresponding author: ]{alireza@apctp.org}
\affiliation{Max-Planck-Institut f\"ur Chemische Physik fester Stoffe, N\"othnitzer Str.~40, 01187 Dresden, Germany}
\affiliation{Asia Pacific Center for Theoretical Physics, Pohang, Gyeongbuk 790-784, Korea}
\affiliation{Department of Physics, POSTECH, Pohang, Gyeongbuk 790-784, Korea}
\affiliation{Max Planck POSTECH Center for Complex Phase Materials, POSTECH, Pohang 790-784, Korea}

\author{S.~E.~Nikitin}
\affiliation{Institut f\"ur Festk\"orper- und Materialphysik, Technische Universit\"at Dresden, 01069 Dresden, Germany}
\affiliation{Max-Planck-Institut f\"ur Chemische Physik fester Stoffe, N\"othnitzer Str.~40, 01187 Dresden, Germany}

\author{A.~S.~Cameron}
\affiliation{Institut f\"ur Festk\"orper- und Materialphysik, Technische Universit\"at Dresden, 01069 Dresden, Germany}

\author{A.\,V.~Dukhnenko}
\altaffiliation{Deceased 22 October 2019.}
\affiliation{I. M. Frantsevich Institute for Problems of Material Sciences of NAS, 3 Krzhyzhanovsky Street, 03680 Kiev, Ukraine}

\author{V.~B.~Filipov}
\affiliation{I. M. Frantsevich Institute for Problems of Material Sciences of NAS, 3 Krzhyzhanovsky Street, 03680 Kiev, Ukraine}

\author{N.~Yu.~Shitsevalova}
\affiliation{I. M. Frantsevich Institute for Problems of Material Sciences of NAS, 3 Krzhyzhanovsky Street, 03680 Kiev, Ukraine}

\author{P.~\v{C}erm\'{a}k}
\affiliation{J\"ulich Center for Neutron Science at MLZ, Forschungszentrum J\"ulich GmbH, Lichtenbergstra{\ss}e 1, 85748 Garching, Germany}
\affiliation{Faculty of Mathematics and Physics, Department of Condensed Matter Physics, Charles University, Ke Karlovu~5, 121\,16, Praha, Czech Republic}

\author{I.~Radelytskyi}
\affiliation{J\"ulich Center for Neutron Science at MLZ, Forschungszentrum J\"ulich GmbH, Lichtenbergstra{\ss}e 1, 85748 Garching, Germany}

\author{A.~Schneidewind}
\affiliation{J\"ulich Center for Neutron Science at MLZ, Forschungszentrum J\"ulich GmbH, Lichtenbergstra{\ss}e 1, 85748 Garching, Germany}

\author{J.~Ollivier}
\affiliation{Institut Laue-Langevin, 71 Avenue des Martyrs CS 20156, 38042 Grenoble Cedex~9, France}

\author{A.~Podlesnyak}
\affiliation{Neutron Scattering Division, Oak Ridge National Laboratory, Oak Ridge, TN~37831, USA}

\author{Z.~Huesges}
\affiliation{\mbox{Helmholtz-Zentrum Berlin f\"{u}r Materialien und Energie GmbH, Hahn-Meitner-Platz 1, 14109 Berlin, Germany}}

\author{J.~Xu}
\affiliation{\mbox{Helmholtz-Zentrum Berlin f\"{u}r Materialien und Energie GmbH, Hahn-Meitner-Platz 1, 14109 Berlin, Germany}}

\author{A.~Ivanov}
\affiliation{Institut Laue-Langevin, 71 Avenue des Martyrs CS 20156, 38042 Grenoble Cedex~9, France}

\author{Y.~Sidis}
\affiliation{Laboratoire L\'{e}on Brillouin, CEA-CNRS, CEA/Saclay, 91191 Gif sur Yvette, France}

\author{S.~Petit}
\affiliation{Laboratoire L\'{e}on Brillouin, CEA-CNRS, CEA/Saclay, 91191 Gif sur Yvette, France}

\author{J.-M.~Mignot}
\affiliation{Laboratoire L\'{e}on Brillouin, CEA-CNRS, CEA/Saclay, 91191 Gif sur Yvette, France}

\author{P.~Thalmeier}
\affiliation{Max-Planck-Institut f\"ur Chemische Physik fester Stoffe, N\"othnitzer Str.~40, 01187 Dresden, Germany}

\author{D.~S.~Inosov}\email[Corresponding author: \vspace{-3pt}]{dmytro.inosov@tu-dresden.de}
\affiliation{Institut f\"ur Festk\"orper- und Materialphysik, Technische Universit\"at Dresden, 01069 Dresden, Germany}

\begin{abstract}\parfillskip=0pt\relax
\noindent
In contrast to magnetic order formed by electrons' dipolar moments, ordering phenomena associated with higher-order multipoles (quadrupoles, octupoles, etc.) are more difficult to characterize because of the limited choice of experimental probes that can distinguish different multipolar moments. The heavy-fermion compound \Ce\ and its La-diluted alloys are among the best-studied realizations of the long-range-ordered multipolar phases, often referred to as ``hidden order''. Previously the hidden order in phase~II was identified as primary antiferroquadrupolar (AFQ) and field-induced octupolar (AFO) order. Here we present a combined experimental and theoretical investigation of collective excitations in the phase~II of \Ce. Inelastic neutron scattering (INS) in fields up to $16.5$~T reveals a new high-energy mode above 14~T in addition to the low-energy magnetic excitations. The experimental dependence of their energy on the magnitude and angle of the applied magnetic field is compared to the results of a multipolar interaction model. The magnetic excitation spectrum in rotating field is calculated within a localized approach using the pseudo-spin presentation for the $\Gamma_8$ states. We show that the rotating-field technique at fixed momentum can complement conventional INS measurements of the dispersion at constant field and holds great promise for identifying the symmetry of multipolar order parameters and the details of inter-multipolar interactions that stabilize hidden-order phases.\vspace{-1pt}
\end{abstract}

\pacs{71.27.+a, 75.30.Mb, 75.40.Gb, 78.70.Nx\vspace{-1pt}}

\maketitle

\vspace{-1pt}\section{Introduction}\vspace{-2pt}
\label{sec:intro}

The high degeneracy and strong Coulomb repulsion of $f\!$-electrons in lanthanide and actinide compounds can lead to exotic quantum matter states at low temperatures \mbox{\cite{thalmeier:05, kusunose:08, kuramoto:09, PaschenLarrea14, paddison:15, ShenLiu19, SibilleGauthier19}}. The hybridization of valence (conduction) electrons with strongly correlated $f\!$-electrons may result in the formation of heavy-fermion metals with quasiparticles that have large effective masses and opening of hybridization gaps that can lead to the Kondo insulator state. Furthermore, at even lower temperatures, broken-symmetry phases (usually magnetic order or superconductivity) may appear driven by residual quasiparticle interactions. Of particular interest are ``hidden order'' (HO) states which consist in spontaneous order of higher rank (rank $r \leq 2l=6$) multipoles of the $f$~electrons $(l=3)$. Here even and odd $r$ correspond to the preservation and breaking of time-reversal symmetry by the order parameter, respectively~\cite{kusunose:08}. In this language, charge or valence (monopole) ordering and spin (dipole) ordering correspond to rank 0 and rank 1 order parameters, respectively. The most well-known examples of materials that host higher-rank order are cubic \Ce\ (rank 2 quadrupole and rank 3 octupole) \cite{shiina:97, CameronFriemel16} and tetragonal \U\ (proposed rank 5 dotriakontapole \cite{ikeda:12, shibauchi:14, thalmeier:14, akbari:15} or rank 4 hexadecapole \cite{KungBaumbach15}). Further examples are the cubic 4$f\!$-skutterudites \cite{kuramoto:09}, 1-2-20 cage compounds \cite{onimaru:16}, and $R_3$Pd$_{20}X_6$ ($R$\,=\,rare earth, $X$\,=\,Si,\,Ge) clathrates \cite{KitagawaTakeda98, NemotoYamaguchi03, GotoWatanabe09, OnoNakano13, PortnichenkoPaschen16}. The symmetry of HO states is difficult to identify because the common x-ray and neutron diffraction (in zero field) yield no signal. Alternative techniques like neutron diffraction in external field \cite{erkelens:87}, resonant x-ray scattering \cite{nagao:01, matsumura:09, nagao:06}, or ultrasonic investigations \cite{nakamura:96, yanagisawa:18} had to be applied to identify combined AFQ and field-induced AFO order in phase~II of \Ce\ below $T_\text{Q}=3.2$~K.

Additional methods for characterizing HO are desirable. One way is to look at the magnetic excitation spectrum in the HO phase which carries the imprint of the multipolar interactions and the HO parameter in its dispersion relations \cite{ShenLiu19, PortnichenkoNikitin19}. In the conventional approach, the dispersion is calculated using a specific candidate model for the HO in some fixed applied field and then compared to the dynamical structure factor measured by single-crystal inelastic neutron scattering (INS). This approach was also tried for \Ce\ where theoretical results, employing extended Holstein-Primakoff (HP) \cite{shiina:03} and random-phase approximation (RPA) methods \cite{thalmeier:98, thalmeier:03, ThalmeierShiina04} have been used to calculate the dispersion and intensity of the magnetic excitation spectrum. Early \cite{bouvet:93} and later more accurate \cite{friemel:12, jang:14} experiments showed that a few modes at high-symmetry points in the simple-cubic Brillouin zone (BZ) could be identified, however an overall determination of the dispersion in the whole BZ still remains elusive. This is not surprising due to several reasons. First of all, to minimize the number of adjustable parameters in the theoretical models, they have been so far restricted to only nearest-neighbor interactions among the multipoles. However, generalized Ruderman-Kittel-Kasuya-Yosida (RKKY) interactions that are mediated by the conduction electrons are expected to be long-range with an oscillatory character in direct space~\mbox{\cite{AkbariThalmeier13, YamadaHanzawa19, HanzawaYamada19}}. There is plenty of experimental evidence that RKKY-type exchange dominates the multipolar interactions in \Ce, as it was shown that the diffuse momentum-space distribution of quasielastic magnetic intensity in the high-temperature paramagnetic state nicely matches the static electronic response (Lindhard) function of the conduction electrons~\cite{KoitzschHeming16, NikitinPortnichenko18}. Recent calculations of the effective RKKY-type exchange terms between different types of multipoles in \Ce~\cite{YamadaHanzawa19, HanzawaYamada19} demonstrate that second-nearest-neighbor contributions are always stronger than the nearest-neighbor ones, and even third-nearest-neighbor terms are not negligible. According to these results, truncating the interactions at $\sim$10\% of the dominant one would still result in a model with at least 5 independent couplings that would have to be treated as adjustable parameters when fitting the experimental magnon dispersions along the conventional approach. To the best of our knowledge, so far no spin-dynamical calculations in the AFQ state that would conform to the complexity of this realistic multi-parameter model have been attempted. The good news, however, is that individual further-neighbor interactions should mainly influence the dispersion away from the zone center, whereas zone-center magnetic excitations are determined by the total sum of all neighbor interactions for each pair of sublattices. For the latter case, only the total sum enters in the Zeeman energy due to the molecular field and in the RPA denominator of the collective magnetic response in Eq.~(\ref{eqn:RPASUS}). Therefore zone-center modes should be much less sensitive to the various individual coupling constants between Ce$^{3+}$ ions. We therefore expect that magnetic excitations at the $\Gamma$ point should show better agreement with the currently available models of multipolar excitations. Moreover, zone-center excitations can be probed not only by INS, but also by electron spin resonance (ESR) \cite{DemishevSemeno06, DemishevSemeno08, DemishevSemeno09, PortnichenkoDemishev16}, hence the experimental results at the $\Gamma$ point can be obtained from two complementary spectroscopic techniques \cite{PortnichenkoDemishev16} and are therefore more reliable than elsewhere in the~BZ.

The second problem in comparing the experimental INS data with the theoretical models consists in the proximity to the antiferromagnetic (AFM) dipolar phase~III that represents the ground state of \Ce\ at low magnetic fields, which is not captured by the previous HP and RPA approaches. Moreover, the intensity of magnetic excitations near high-symmetry $\Gamma$ and $R$ points in the simple-cubic BZ is enhanced by conduction-$4f$ electron Kondo screening, which leads to itinerant heavy $4f$ quasiparticles \cite{TazaiKontani19} and ultimately to spin-exciton formation within the AFM phase \cite{friemel:12, akbari:12}. This possibility is not contained in the theory based on completely localized $4f\!$-electron picture, hence a reliable comparison of experimental data with the present models is only possible at high magnetic fields deep within phase~II, where the AFM order parameter is suppressed.

The third difficulty is that the calculated inelastic magnetic response in phase~II is limited to the dipolar response function, whereas strictly speaking neutron scattering is sensitive to all odd-rank \textit{magnetic} multipoles (such as dipole, octupole, dotriakontapole etc.). At short scattering vectors, $|\mathbf{Q}|\rightarrow0$, the form factor is expected to suppress all higher-order multipolar contributions, which justifies the so-called \textit{dipolar approximation} that has been employed for calculating the dynamical structure factor even for multipolar-ordered systems. The non-monotonic form factor of higher-order multipoles vanishes at $\mathbf{Q}=0$ and then starts to increase until reaching a maximum at some finite momentum transfer. This has been established from both theory and experiment for elastic neutron scattering \cite{ShiinaSakai07, KuwaharaIwasa07, kuramoto:09, Shiina12}, but to the best of our knowledge, the highly involved theory of INS beyond the dipolar approximation \cite{BalcarLovesey89, JensenMackintosh91} was never successfully applied to calculate the dynamical response functions for any compound with a multipolar-ordered ground state. On the other hand, experiments clearly demonstrate that the intensity of the $\Gamma$-point excitation in \Ce\ increases with wave vector when going from the shortest $\Gamma'(001)$ to the second-shortest $\Gamma''(110)$ reciprocal-lattice vector \cite{jang:14}, suggesting a non-monotonic form factor characteristic of multipolar moments that cannot be described in the dipolar approximation. Access to the same excitation at even longer wave vectors is restricted by the worsening of energy resolution imposed by the kinematic constraints, and therefore the actual share of non-dipolar spectral weight in the experimental INS spectrum remains unknown. We expect that it should change with the applied magnetic field, as it induces secondary dipolar and octupolar order parameters on top of the primary AFQ order, thereby activating additional magnetic degrees of freedom that are subject to collective fluctuations. The multipolar corrections should increasingly influence the measured intensities of magnetic excitations towards higher $|\mathbf{Q}|$, leading to deviations from the established spin-dynamical models.

In high magnetic fields, sufficiently far above phase~III, when the Kondo and spin-exciton effects due to itineracy of heavy $f\!$-electrons are suppressed, the localized approach becomes appropriate. In particular, the gapped Goldstone mode around the $\Gamma$ point, which was predicted by the theory \cite{thalmeier:03, ThalmeierShiina04}, has been eventually observed in an INS experiment~\cite{jang:14}, in quantitative agreement with the ferromagnetic resonance seen earlier using ESR spectroscopy \cite{DemishevSemeno06, DemishevSemeno08, DemishevSemeno09, PortnichenkoDemishev16}. However, due to the difficulties outlined above, we found no quantitative agreement between the theory and the observed dispersion of multipolar excitations throughout the BZ. In the present work, we propose an alternative approach to analyze the fingerprints of HO in the magnetic excitation spectrum, which appears to be more promising in providing quantitative information while staying within the same theoretical framework. Rather than changing continuously the momentum transfer and following the dispersion of multipolar excitations in momentum space, one may fix the wave vector at some high-symmetry point with large INS intensity (e.g. $R$ or $\Gamma$) and instead change the direction and strength of the applied magnetic field. Ideally, the field could be continuously rotated, so that changing the field alone would give access to a three-dimensional parameter space for every single value of the momentum-transfer vector $\mathbf{Q}$. However, due to the limited measurement time, we were practically restricted only to several high-symmetry directions of the magnetic field in this work.

We demonstrate that not only the energy and intensity, but even the number of experimentally observable multipolar excitations changes as a function of field direction, offering an abundant source of experimental information that remained virtually unexplored in all the previous measurements. Thermodynamically, the anisotropy of the critical fields of phase~II in CeB$_6$ was found to be determined by the underlying AFQ/AFO hidden order. Therefore, a similar anisotropy in the pattern of collective modes under the field rotation is not at all surprising and can be calculated within the framework of available models. We will demonstrate the usefulness of this new approach to HO symmetry on the canonical example of \Ce\ multipole model, where the order parameter is already known from other methods. Our work has the twofold aim of proposing a general new method for HO investigation with INS for fixed momentum transfer in a rotating field and making concrete experimental and theoretical analysis of magnetic excitation modes in \Ce\ as a function of field direction and magnitude, in particular as guidance for further experiments and future theoretical developments.

The paper is organized as follows. In Sec.~\ref{Sec:Experiment} we describe new INS experiments on \Ce\ obtained in fields up to 16.5~T. We present data measured in several high-symmetry points of the BZ for different field directions, obtained in multiple INS experiments on the same single-crystalline sample of \Ce. Then, in Secs.~\ref{sec:model} and \ref{sec:RPA} we recapitulate the basic RPA-type response theory for calculating the magnetic excitation spectrum for a localized $\Gamma_8$ model in the presence of quadrupolar and induced octupolar HO. In Sec.~\ref{Sec:TheoryDispersion} we present the results for the multipolar mode dispersions using composite $\Gamma^+_5/\Gamma^-_2$ HO model of \Ce\ as visible in the intensity of the dipolar structure factor. In Sec.~\ref{sec:field} we present analogous calculations in dependence on the field strength and direction in the BZ high-symmetry points. We then give a detailed comparison of the theoretical and experimental results, showing clear evidence for a new high-energy branch of multipolar excitations observed for the first time in \Ce. Finally, we will summarize our results in Sec.~\ref{sec:summary}, where we argue that our results on \Ce\ serve as a proof of principle for a new alternative way of obtaining quantitative information about hidden-order symmetry and interactions between multipole degrees of freedom in $f$-electron materials with non-dipolar order parameters. This method should extend the conventional approach to the analysis of neutron-scattering data and may find applications in a much broader class of correlated-electron systems.

\vspace{-2pt}\section{Experimental setup and results}\vspace{-3pt}\label{Sec:Experiment}

\subsection{Sample description and experimental configurations}\vspace{-2pt}

Most of the experimental results presented in this paper were measured on the same rod-shaped Ce$^{11}$B$_{6}$ single crystal with a mass of $\sim$4\,g, which was previously used for experiments reported in Refs.~\cite{friemel:12, jang:14, PortnichenkoDemishev16}. To minimize neutron absorption by the highly absorbing $^{10}$B isotope present with a 19.9\% abundance in natural boron, the crystal was grown by the floating-zone method from the isotope-enriched $^{11}$B powder (Ceradyne Inc., 99.6\,\% enrichment level), as described elsewhere \cite[see Methods]{friemel:12}. The growth axis of the crystalline rod was approximately parallel to the [110] crystallographic axis, with a deviation of a few degrees. Before every experiment, the crystal was pre-aligned in the corresponding scattering plane (which varied from one experiment to another) using an x-ray Laue camera.

We have measured the magnetic excitation spectrum of \Ce\ by neutron spectroscopy for different field directions using 6 different instruments. These are the cold-neutron triple-axis spectrometers (TAS) PANDA (Heinz Maier-Leibnitz Zentrum, Garching, Germany)~\cite{schneidewind:06}, FLEXX (Helmholtz-Zentrum Berlin, Germany)~\cite{LeCastro13, habicht:15}, ThALES (Institut Laue-Langevin, Grenoble, France)~\cite{boehm:08, ILLData4-01-1537, ILLData4-01-1583, ILLData4-01-1624}, and 4F2 (Laboratoire L\'eon Brillouin, Saclay, France), as well as the disk chopper time-of-flight (TOF) spectrometer IN5 (Institut Laue-Langevin, Grenoble)~\cite{ollivier:11, ILLData4-03-1710} and the cold-neutron chopper spectrometer CNCS (Spallation Neutron Source, Oak Ridge National Laboratory, USA) \cite{ehlers:11, EhlersPodlesnyak16}. The latter were required to measure the magnetic excitation spectrum at a constant magnetic field over the entire BZ. The corresponding experimental configurations and sample orientations are summarized in Table~\ref{Tab:Setups}.

\begin{table}[t!]
\begin{center}\vspace{4pt}
\begin{tabular}{@{}l|l|l|l@{\hspace{-8pt}}l|l@{}}
   \toprule
     No. & Instrument & Neutron & Magnet & & Field \\
       &      & energy &    & & direct.\\
   \hline
     1 & FLEXX\,@\,HZB & $E_{\rm f}=3.50$~meV & 15~T   & (VM-1B) & $[1\overline{1}0]$  \\
     2 &        & $E_{\rm f}=4.66$~meV & ~~ '' ~~ & with booster & $[1\overline{1}0]$  \\
   \hline
     3 & ThALES\,@\,ILL & $E_{\rm f}=3.50$~meV & 15~T & (158OXHV26) & $[1\overline{1}0]$  \\
     4 &        & $E_{\rm f}=3.50$~meV & 10~T & (256CRHV41) & $[001]$  \\
     5 &        & $E_{\rm f}=3.50$~meV & 15~T & (158OXHV26) & $[1\overline{1}1]$  \\
   \hline
     6 & PANDA\,@\,MLZ & $E_{\rm f}=3.50$~meV & 7.5~T      & & $[1\overline{1}0]$  \\
      &        & $E_{\rm f}=4.66$~meV & 7.5~T      & & $[1\overline{1}0]$  \\
     7 &        & $E_{\rm f}=3.50$~meV & 8~T       & & $[1\overline{1}1]$  \\
     8 &        & $E_{\rm f}=3.24$~meV & 5~T       & & $[001]$  \\
   \hline
     9 & 4F2\,@\,LLB  & $E_{\rm f}=3.50$~meV & 9~T       & & $[1\overline{1}2]$  \\
   \hline
     10 & IN5\,@\,ILL  & $E_{\rm i}=3.27$~meV & 2.5~T & (90ILHV42) & $[1\overline{1}0]$  \\
     11 & CNCS\,@\,ORNL & $E_{\rm i}=3.15$~meV & 5~T       & & $[1\overline{1}0]$  \\
   \bottomrule
\end{tabular}
\caption{Experimental INS setups used in the present work. TAS instruments are listed as setups 1--9, and TOF spectrometers as setups 10 and 11. The field direction in all setups is vertical, so that the scattering plane is always orthogonal to the direction given in the last column. Whenever different magnets with the same maximal field were available at a given instrument, the specific model of the magnet is given in brackets to avoid ambiguity.}
\label{Tab:Setups}
\end{center}\vspace{-1.6em}
\end{table}

\begin{figure}[b!]
\includegraphics[width=\linewidth]{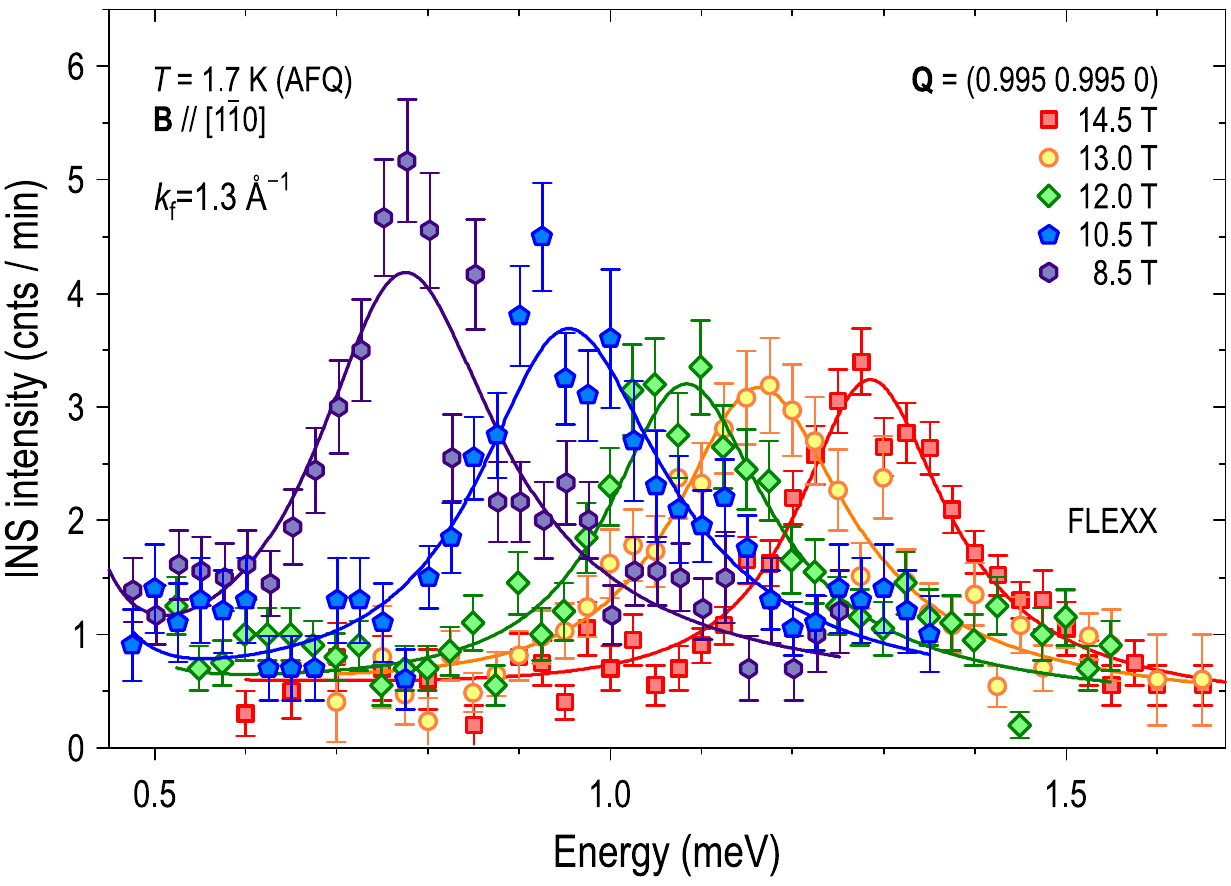}
\caption{Unprocessed INS spectra measured near the zone center $\Gamma''(110)$ for $\mathbf{B} \parallel [1\overline{1}0]$ using setup~1.\vspace{-4pt}}
\label{Fig:GammaRawScan110}
\end{figure}

\begin{figure*}[!t]
\includegraphics[width=0.8\textwidth]{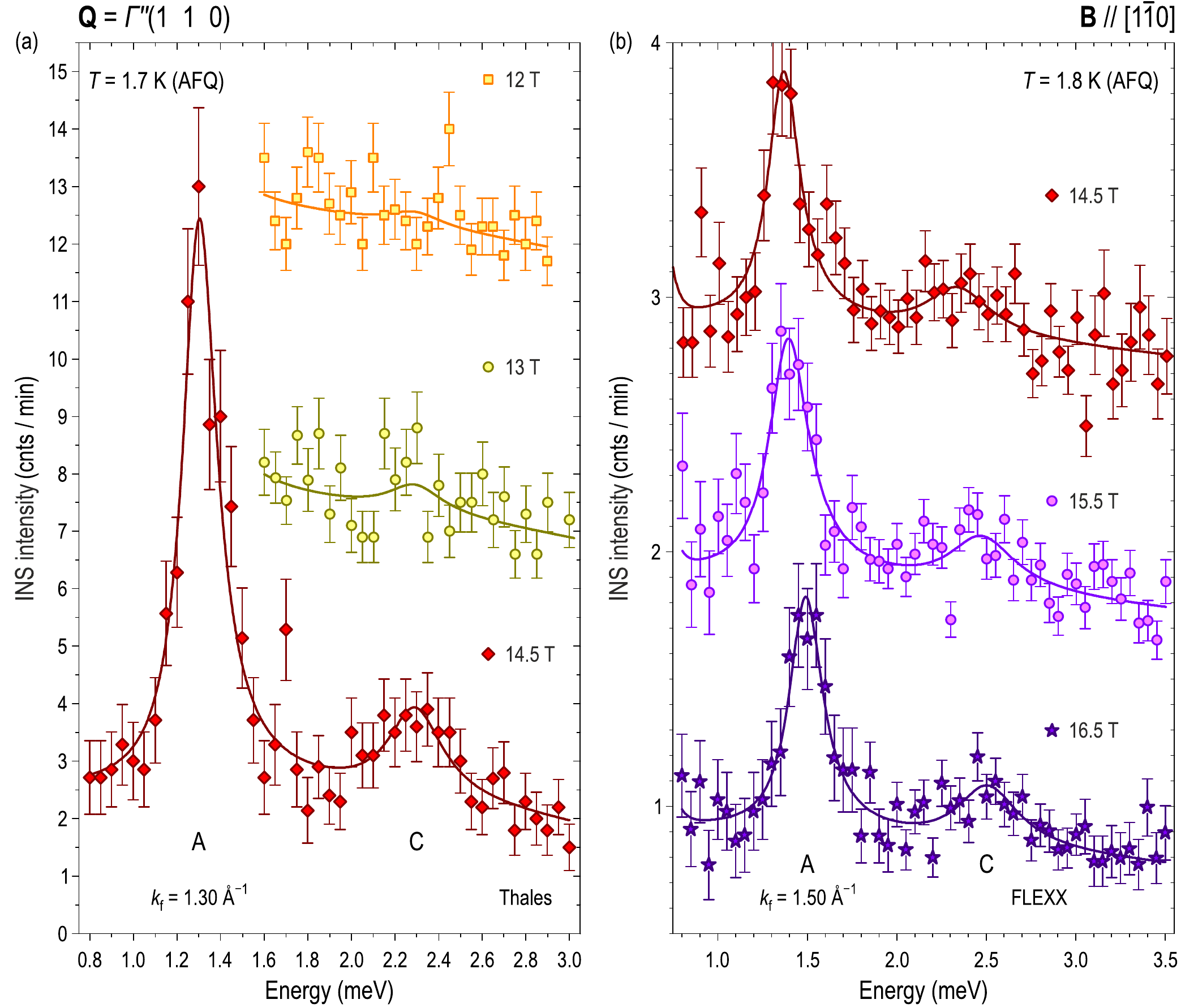}
\caption{Unprocessed INS spectra measured near the zone center $\Gamma''(110)$ for $\mathbf{B}\parallel[1\overline{1}0]$ using setups 3~(left) and 2~(right). The data are offset vertically for clarity. They illustrate the appearance of resonance~C (smaller peak) at a higher energy with respect to resonance~A (stronger peak) at magnetic fields above 14~T. Both peak positions shift upwards in energy with increasing field. Solid lines are Lorentzian~fits.}
\label{Fig:ThalesFLEXXbooster110}
\end{figure*}

In every TAS experiment, a cold Be filter was used in order to avoid higher-order neutron contamination from the monochromator, and the final wave vector of the neutrons was fixed to $k_{\rm f}=1.3$\,\AA$^{-1}$ ($E_{\rm f}=3.50$~meV) or $1.5$\,\AA$^{-1}$ (4.66~meV), as indicated for every dataset. TOF measurements were done with the incident neutron wavelength fixed at $\lambda_{\rm i}=5$\,\AA\ ($E_\text{i}=3.27$~meV) for IN5 and 5.1\,\AA\ (3.15~meV) for CNCS experiments. The magnetic field was applied in the vertical direction (perpendicular to the scattering plane) using various cryomagnets available at the corresponding neutron facilities. For PANDA experiments, we used either the 7.5~T liquid-free vertical-field cryomagnet with a $^{3}$He insert or the 5\,T and 8\,T LHe-cooled vertical-field cryomagnets for measurements with magnetic field along the $[1\overline{1}0]$, $[001]$ and $[1\overline{1}1]$ crystal directions, respectively. For the experiment at 4F2 we used a vertical-field 9~T magnet. In the FLEXX and ThALES experiments, we used similar 15~T magnets equipped with a lambda point refrigerator. To achieve the maximal field of 16.5~T, the same magnet was used with an additional dysprosium ``booster'', which concentrates magnetic field lines in a smaller sample volume, so that higher magnetic field can be achieved at the expense of the smaller sample size. For this unique configuration, a smaller piece of an identical crystal rod ($\sim$\,0.9~g in mass) had to be cut to fit into the sample space of the ``booster''. In the IN5 and CNCS experiments, we used the low-background 2.5~T ``orange''-type cryostat based magnet and the 5~T cryomagnet, respectively.

\vspace{-12pt}\subsection{Magnetic field along $[1\overline{1}0]$}\vspace{-5pt}
\label{Sec:INS110}

\subsubsection{Zone-center excitations}\vspace{-5pt}

We start the presentation of our results with the high-field INS data obtained with the magnetic field $\mathbf{B}\parallel[1\overline{1}0]$, because this field direction has been most extensively studied in the past, and in particular in our earlier work \cite{PortnichenkoDemishev16} we already presented both the field dependence for fields up to 7~T and the momentum dependence of multipolar excitations for certain values of the field in this direction. We found perfect quantitative agreement between the energy of the single resonance at the $\Gamma$ point and the ferromagnetic resonance observed in ESR (resonance~A) within the measured field range. At the same time, a second ESR resonance (resonance~B) was seen at fields above 12~T \cite{DemishevSemeno08}, which we did not cover in our initial experiment. We therefore decided to extend our INS measurements up to higher fields, expecting to find the emergence of the resonance B at lower energies relative to the main resonance, as suggested by the ESR data. However, the corresponding energy scans measured with setup~1 at the base temperature of 1.7~K, which are presented in Fig.~\ref{Fig:GammaRawScan110}, show no signatures of such an additional peak in the expected energy range of 0.9--1.1~meV up to the maximal field of 14.5~T accessible with this particular experimental setup. The background in our range of interest is clean and essentially constant, and the energy resolution is sufficiently narrow so that the tail of the peak corresponding to resonance~A does not overlap with the expected position of resonance~B at such high magnetic fields. For instance, at 14.5~T the expected peak splitting is 0.35~meV, while the instrument resolution is less than 0.1~meV \cite{LeCastro13}. Therefore, we can conclude that resonance~B is not visible to neutrons, possibly due to certain selection rules that are different from those of ESR.

\begin{figure}[t!]
\includegraphics[width=\linewidth]{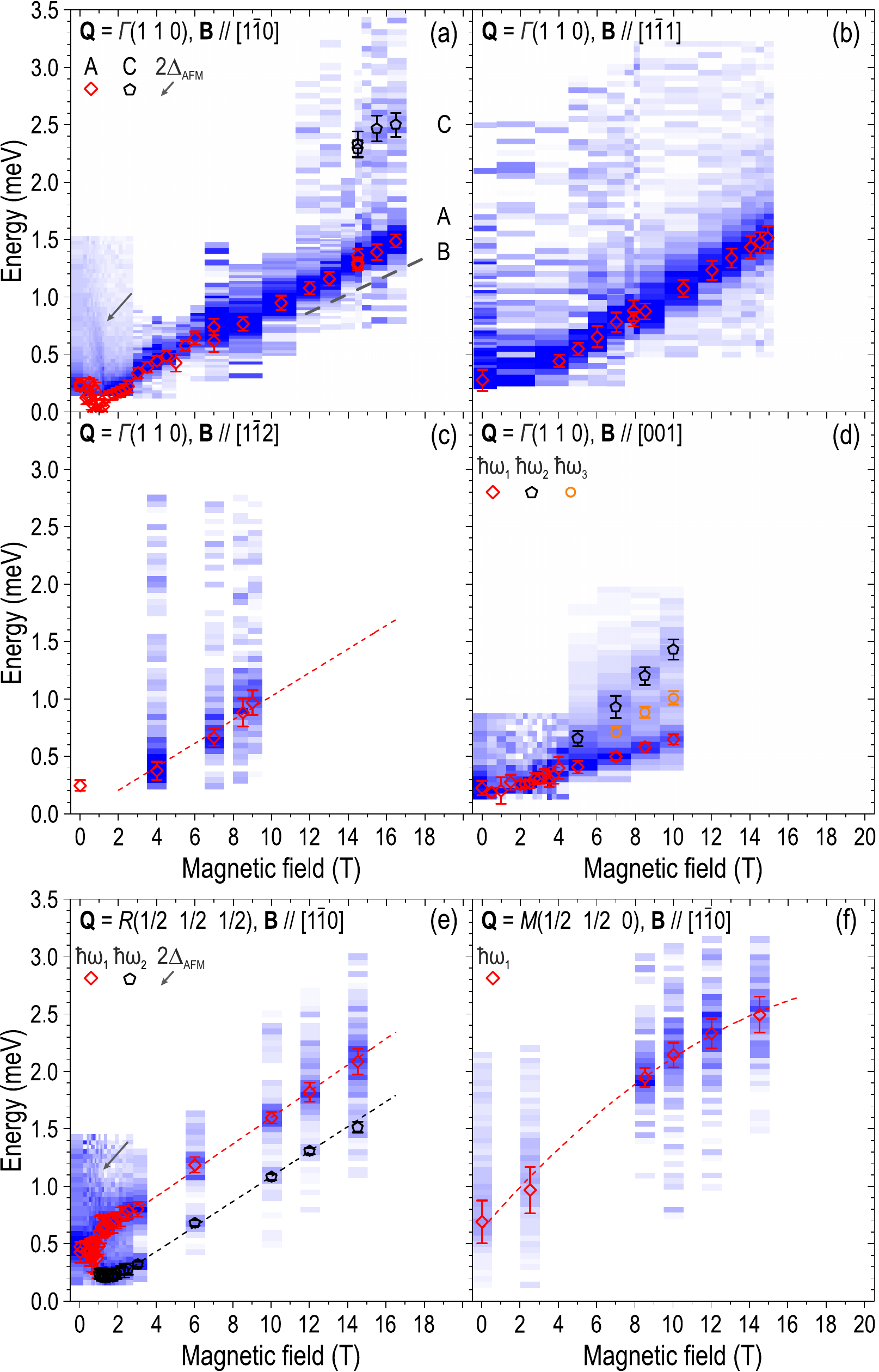}\vspace{1pt}
\caption{Summary of the neutron spectra, plotted as a function of magnetic field: (a--d)~at the $\Gamma''(110)$ point for different magnetic field directions; (e,\,f) at the $R(\frac{1}{2}\frac{1}{2}\frac{1}{2})$ and $M(\frac{1}{2}\frac{1}{2}0)$ points for $\mathbf{B} \parallel [1\overline{1}0]$. Darker blue color corresponds to higher INS intensity. Data points mark fitted peak positions. The dashed line in panel (a) marks the expected position of resonance~B seen in ESR data (see text). The resonances A and C observed in INS experiments are also indicated. We do not extend this notation to other field directions, because the correspondence between these resonances and the three resonances $\omega_1$, $\omega_2$, $\omega_3$ for $\mathbf{B}\parallel[001]$ in (d) has not yet been clarified. Dotted lines in panels (c),\,(e),\,(f) are guides to the eyes.\vspace{-1pt}}
\label{Fig:SummaryField}
\end{figure}

To confirm this unexpected result, we repeated the experiment at the high-flux spectrometer ThALES (setup~3)~\cite{ILLData4-01-1537}, the corresponding data are presented in Fig.~\ref{Fig:ThalesFLEXXbooster110}\,(a). The absence of the peak associated with resonance~B was reproduced, but extending the spectrum to higher energies revealed a new peak at a higher energy of $\sim$2.25~meV in the 14.5~T dataset that rapidly vanished when the field was decreased below its maximal value. This result suggests that an additional resonance (resonance~C) forms above $\sim$14~T, at approximately the same field as resonance~B, but on the opposite side of the intense peak corresponding to resonance A. Because 14.5~T was the highest field available on ThALES, we performed one more experiment at the FLEXX spectrometer using the 15~T magnet equipped with a dysprosium ``booster'' (setup~2), which has a much smaller sample volume, so the sample mass and, consequently, the count rate were reduced (which was partly compensated by choosing a larger value of $k_{\rm f}=1.5$~\AA$^{-1}$). The resulting spectra, shown in Fig.~\ref{Fig:ThalesFLEXXbooster110}\,(b), had to be counted up to 2~h per data point. In this configuration, we could reach fields up to 16.5~T, following a shift in the energy of resonance~C with approximately the same slope ($g$-factor) as that of the more intense resonance~A. This leaves no doubt about the magnetic origin of the second peak.

In Fig.~\ref{Fig:SummaryField}\,(a), we summarize these new data (setups \mbox{1--3}) together with the previously published results for the same field direction (setups 6,\,10,\,11) \cite{PortnichenkoDemishev16} in the form of a color map. The fitted positions of resonances~A and C are shown with red diamonds and black pentagon symbols, respectively. One can see that the energy of both resonances shifts upwards with increasing field with approximately the same slope. The expected position of resonance~B, according to ESR results \cite{DemishevSemeno08}, is shown with a dashed gray line. In the low-field region within Phase~III, an arrow marks the previously reported feature at twice the AFM charge gap \cite{FriemelJang15}.

\begin{figure}[b!]\vspace{-2pt}
\includegraphics[width=\linewidth]{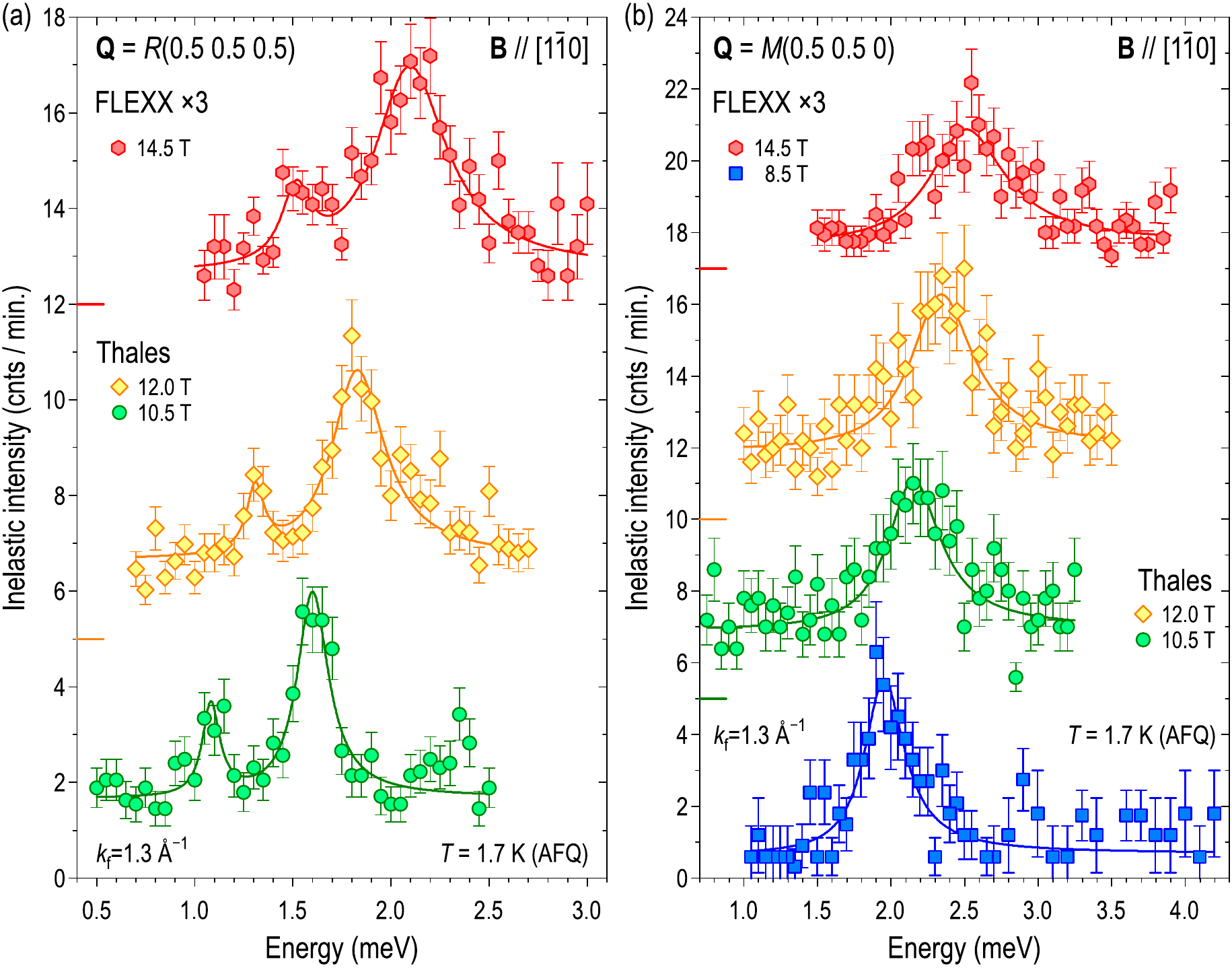}\vspace{-1pt}
\caption{Unprocessed INS data, measured at the $R(\frac{1}{2}\frac{1}{2}\frac{1}{2})$ and $M(\frac{1}{2}\frac{1}{2}0)$ points in different magnetic fields using setups 1 and 3. The curves are shifted vertically for clarity by an amount indicated on the vertical axis with horizontal bars of corresponding colors. Solid lines are fits with Lorentzian peak functions on top of a background. The data measured with setup 1 were additionally multiplied by 3 to obtain similar peak intensity in all curves.\vspace{-3.5pt}}
\label{Fig:ComparisonRandM}
\end{figure}

\vspace{-6pt}\subsubsection{Dispersion of field-induced multipolar excitations}\vspace{-5pt}
\label{SubSec:Dispersion}

\begin{figure}[t!]\vspace{2pt}
\includegraphics[width=\linewidth]{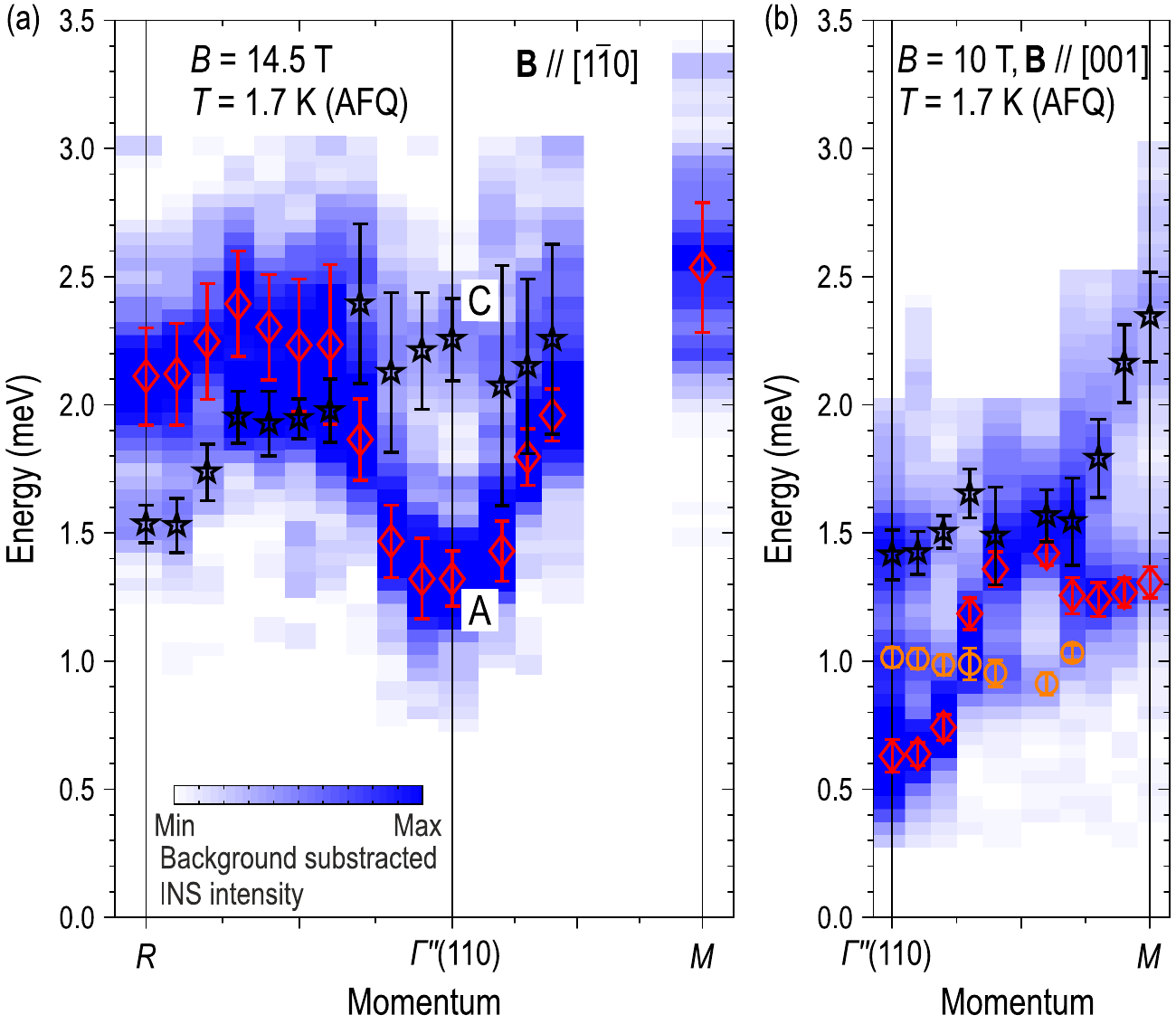}
\caption{INS intensity vs.~energy and momentum along straight segments connecting the $R(\frac{1}{2}\frac{1}{2}\frac{1}{2})$, $\Gamma''(110)$, and $M(\frac{1}{2}\frac{1}{2}0)$ points: (a)~in a magnetic field of 14.5\,T, applied along the $[1\overline{1}0]$ direction using setups 1 and 3; (b)~in a magnetic field of 10\,T, applied along the $[001]$ direction using setup~4. The color map shows raw data after background subtraction and subsequent smoothing with a one-dimensional Gaussian filter characterized by a FWHM of 0.1\,meV (to reduce statistical noise). Resonances A and C at the $\Gamma$ point are marked as such. Markers show the fitted peak positions.}
\label{Fig:DispersionField}
\end{figure}

In the two bottom panels of Fig.~\ref{Fig:SummaryField}, similar field dependencies for the same magnetic field direction $\mathbf{B}\parallel[1\overline{1}0]$ are presented for two other high-symmetry points in the BZ. The data at $R(\frac{1}{2}\frac{1}{2}\frac{1}{2})$ in Fig.~\ref{Fig:SummaryField}\,(e) combine our previously published results for fields below 6~T (setups~6 and 10 from Refs.~\cite{FriemelJang15} and \cite{PortnichenkoDemishev16}, respectively) with new measurements using setups 1 and 3 that extend to higher fields up to 14.5~T. The corresponding raw data for the three highest field values are also presented in Fig.~\ref{Fig:ComparisonRandM}\,(a). One can see that the two excitations that were observed earlier in phase~II \cite{PortnichenkoDemishev16} persist also in higher fields, following the same linear trends with an approximately equal slope. The data at $M(\frac{1}{2}\frac{1}{2}0)$ in Fig.~\ref{Fig:SummaryField}\,(f) consist of the zero-field and 2.5~T spectra from Ref.~\cite{PortnichenkoDemishev16} (setup~10) \cite{ILLData4-03-1710} and four new scans at higher fields between 8.5 and 14.5~T, which are shown separately in Fig.~\ref{Fig:ComparisonRandM}\,(b). They indicate the presence of at least one excitation whose energy increases monotonically with the magnetic field.

The momentum dependence of multipolar excitations at intermediate wave vectors is best revealed in Fig.~\ref{Fig:DispersionField}\,(a), where the INS intensity has been mapped out using setups~1 and 3 as a function of momentum along the polygonal path $R(\frac{1}{2}\frac{1}{2}\frac{1}{2})$\,--\,$\Gamma''(110)$\,--\,$M(\frac{1}{2}\frac{1}{2}0)$ in momentum space. These data were measured with the maximal field of 14.5~T applied along $[1\overline{1}0]$. One can see that the more intense higher-energy excitation at the $R$ point appears to be continuously connected to resonance~A at the zone center, whereas the weaker low-energy resonance at $R$ seems to cross this branch and re-emerge as resonance~C at the $\Gamma$ point. At the same time, both resonances approach each other along the $\Gamma M$ line, so that only a single peak is observed at the $M$ point. However, from our earlier results \cite{PortnichenkoDemishev16}, we know that the low-energy peak at the $R$ point emerges already at 1.7~T, simultaneously with phase~II. On the other hand, resonance~C at the $\Gamma$ point is only visible above 14~T, which leaves an open question about the detailed evolution of this branch along the $R\Gamma$ line with increasing field. Unfortunately, the two excitations cannot be clearly resolved close to their crossing point between $R$ and $\Gamma$, hence it is unlikely that similar momentum-dependent maps at intermediate magnetic fields will help to answer this question conclusively.

Finally, it is useful to compare the dispersion of multipolar excitations in Fig.~\ref{Fig:DispersionField}\,(a) with similar momentum-resolved intensity maps measured in smaller magnetic fields of 2.5 and 5~T, which were presented earlier~\cite{PortnichenkoDemishev16}. Across the whole field range, the bottom of the magnon band is located at the $\Gamma$ point, and the maximum of the dispersion is reached at the $M$ point. The total magnon bandwidth does not change and stays at approximately 1.2~meV in the whole field range, while the whole spectrum shifts upwards in energy with increasing field. At 2.5~T, there are two resonances at the $R$ point, of which the lower one displays a steep upward dispersion along the $R\Gamma$ line, ultimately merging with the more intense branch emanating from the zone center. As we can see now, these two branches actually cross each other, leading to the appearance of resonance C at the $\Gamma$ point that can only be seen in high magnetic fields. This behavior is qualitatively consistent with the theoretical predictions that will be presented in Sec.~\ref{Sec:TheoryDispersion}.

\vspace{-3pt}\subsection{Anisotropy with respect to the field direction}\vspace{-3pt}
\label{Sec:ExpAnisotropy}

\begin{figure*}[t!]
\includegraphics[width=\linewidth]{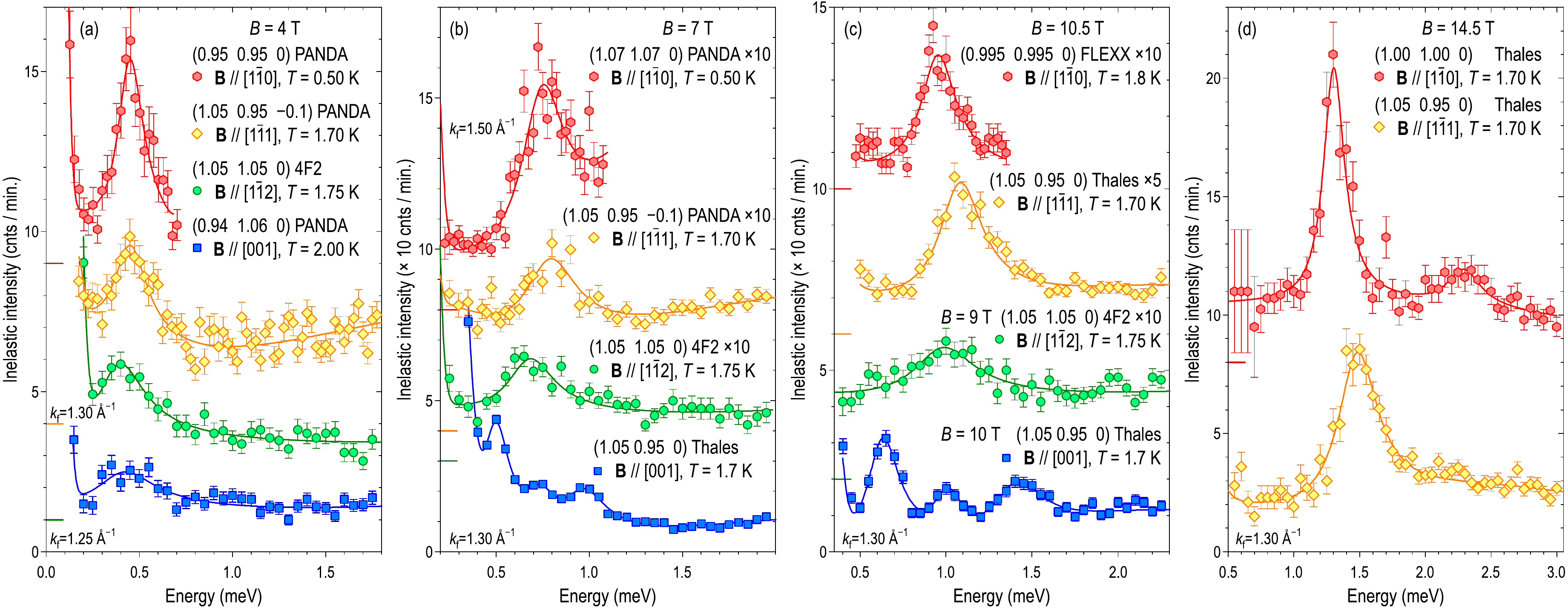}
\caption{Comparison of the raw INS spectra at the $\Gamma''(110)$ point (or very close to it, to avoid Bragg contamination), measured in magnetic fields applied along different crystallographic directions. The reciprocal-space vector, measurement temperature, magnetic field and its direction are indicated in the legends. Note that the data were measured using different spectrometers in different configurations, therefore the absolute intensity is not directly comparable. The curves are shifted vertically for clarity by an amount indicated on the vertical axis with horizontal bars of corresponding colors. Solid lines are empirical fits that serve as guides to the eyes.\vspace{-3.5pt}}
\label{Fig:ComparisonFields}
\end{figure*}

\begin{figure}[b!]\vspace{-21pt}
\includegraphics[width=0.95\linewidth]{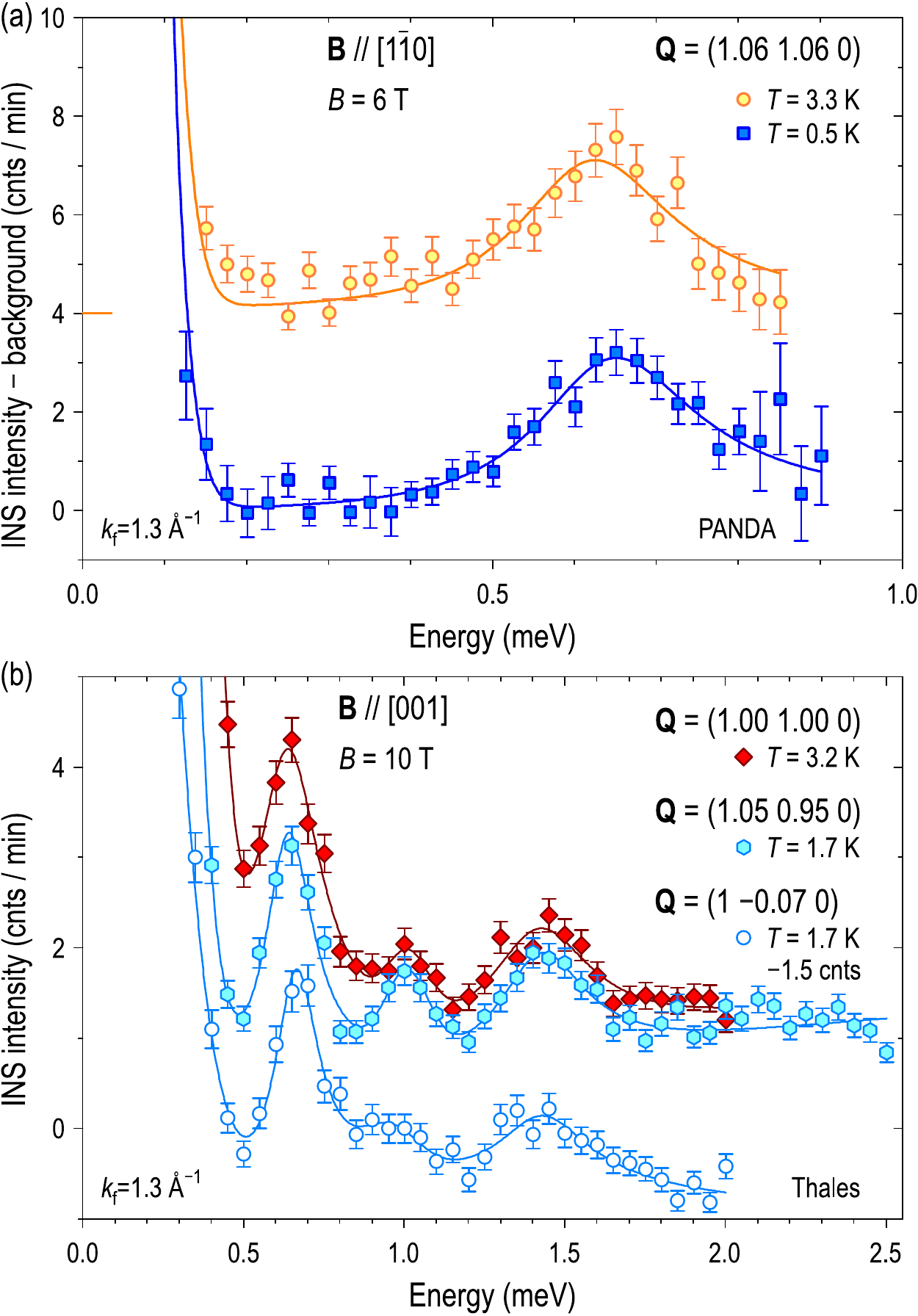}\vspace{-1pt}
\caption{Comparison of the INS data collected in the vicinity of the $\Gamma$ point at the base temperature and at an elevated temperature, as indicated in the legend. The panels (a) and (b) were measured at 6 and 10~T using setups 6 and 4, respectively. The bottom curve in panel~(b), which is shifted down by --1.5 units for clarity, shows an equivalent spectrum measured near the $\Gamma'(100)$ point to confirm the reproducibility of all three peaks. Note a small difference in $\mathbf{Q}$ between the two top spectra in the bottom panel, which explains a difference in nonmagnetic background, while the position of all three peaks remains unchanged.}
\label{Fig:TempIndep}
\end{figure}

We proceed by demonstrating the dependence of the INS spectra on the direction of the applied magnetic field. Figure~\ref{Fig:ComparisonFields} shows the comparison of raw INS data measured in the vicinity of the $\Gamma''(110)$ point at the same or similar values of magnetic field, but for different field orientations. Additional spectra are also presented in Figs.~\ref{Fig:GammaRawScans001} and \ref{Fig:GammaRawScans112and111} of the Supplemental Material \cite{SupplementalMaterial}. At a relatively low field of 4~T [Fig.~\ref{Fig:ComparisonFields}\,(a)], the spectra look nearly independent of the field direction, except for the difference in absolute intensity and the signal-to-background ratio, which can be explained by different experimental setups and by the different orientation of the crystal rod with respect to the beam~\footnote{Our single-crystal sample represents a 40-mm-long cylindrical rod $\sim$6~mm in diameter, with its axis nearly parallel to the $[1\overline{1}0]$ crystal direction. This sample geometry is optimal for measurements with $\mathbf{B}\parallel[1\overline{1}0]$ for two reasons. First, the homogeneity of the magnetic field is better along the magnet axis than perpendicular to it, hence a tilted orientation of the sample would lead to a less homogeneous field distribution along the rod, causing an additional broadening of the peak. Second, even the small concentration ($\lesssim$\,0.4\%) of the highly absorbing $^{10}$B isotope in our isotope-enriched sample causes considerable absorption of cold neutrons. The absorption increases as the sample is tilted away from its vertical orientation, leading to a decrease in the signal intensity. As a result, measurements with $\mathbf{B}\parallel[001]$, which require the largest tilting angle of 45$^\circ$, typically show weaker signals as compared to other field orientations.}. At~higher magnetic fields [Fig.~\ref{Fig:ComparisonFields}\,(b--d)], the differences become more pronounced. This is seen not only in the different energy of the resonance as a function of field direction, but also in the appearance of additional spectral lines for $\mathbf{B}\parallel[001]$. For this field orientation, one can see three peaks that shift as a function of field with a different slope, as shown in Fig.~\ref{Fig:GammaRawScans001} in the Supplemental Material \cite{SupplementalMaterial} and in Fig.~\ref{Fig:SummaryField}\,(d). The same three peaks are also reproduced near the equivalent zone-center point $\Gamma'(100)$, as shown in the bottom curve of Fig.~\ref{Fig:TempIndep}\,(b). This example clearly demonstrates that INS measurements in a rotating field offer additional information about collective modes in CeB$_6$ that have so far avoided direct observations in either INS or ESR experiments. Qualitative differences are also seen at other points in the BZ. For instance, the dispersion along the $\Gamma M$ direction, measured using setup~4 \cite{ILLData4-01-1624}, is depicted in Fig.~\ref{Fig:DispersionField}\,(b). An additional peak is observed at the $M$ point, which appears to correspond to two excitations that split closer to the zone center.

The measurements presented so far were conducted using different experimental setups with different cryomagnets and at somewhat different temperatures, depending on the base temperature of the available sample environment. These temperatures were always much lower than the ordering temperature of phase~II, and therefore one expects that these small temperature differences can be neglected. Nevertheless, for a reliable comparison of the data acquired with different setups, we had to ensure that the possible temperature dependence of the excitation energies is not detectable within our experimental accuracy. This is demonstrated in Fig.~\ref{Fig:TempIndep}, where measurements at the base temperature of 0.5 and 1.7~K in setups 6 and 4, respectively, are compared with similar spectra measured at an elevated temperature of 3.3 or 3.2~K. For both field directions, $\mathbf{B}\parallel[1\overline{1}0]$ and $\mathbf{B}\parallel[001]$, the spectrum shows no detectable changes with temperature. In Fig.~\ref{Fig:TempIndep}\,(b), the higher background of the higher-temperature curve is a consequence of non-magnetic scattering, as this spectrum was measured exactly at the allowed Bragg peak position, while the low-temperature data were measured with a small offset in $\mathbf{Q}$. Nevertheless, the positions of all three magnetic peaks in these spectra are identical within the resolution of our experiment. This way, we can exclude the effect of small temperature variations on the outcome of our measurements.

\begin{figure}[t!]\vspace{-2pt}
\includegraphics[width=0.795\linewidth]{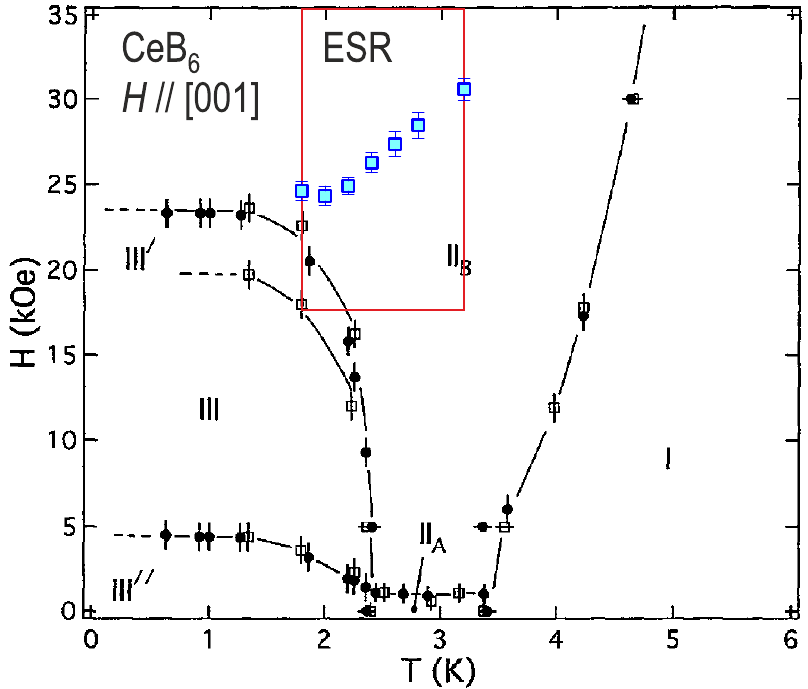}\vspace{-1pt}
\caption{The phase diagram of CeB$_6$ for the magnetic field parallel to $[001]$, reproduced from Ref.\,\citenum{NakamuraGoto95}. The red rectangle shows the part of the phase diagram covered by ESR measurements of the temperature-dependent $g$-factor by Semeno \textit{et al.}~\cite{SemenoGilmanov16, SemenoGilmanov17}. Blue squares mark the positions of the 60~GHz resonance at different temperatures.\vspace{-5pt}}
\label{Fig:PhaseDiagESR}
\end{figure}

At first glance, this conclusion may seem to be in contradiction with the recent ESR results by Semeno \textit{et al.}~\cite{SemenoGilmanov16, SemenoGilmanov17, Gilmanov19}. In this experiment, the energy of the zone-center magnetic resonance has been measured for different directions of the magnetic field between $\mathbf{B}\parallel[110]$ and $\mathbf{B}\parallel[001]$, revealing a significant ($\sim$10\%) anisotropy in the $g$-factor with a strong temperature dependence, which was most pronounced for $\mathbf{B}\parallel[001]$. The $g$-factor measured in ESR spectroscopy can be directly related to the slope of the linear field dependence of magnetic excitations at the $\Gamma$ point, and we have shown \cite{PortnichenkoDemishev16} that the values obtained from ESR and INS measurements for the field direction $\mathbf{B}\parallel[1\overline{1}0]$ are in perfect agreement with each other. On the other hand, for $\mathbf{B}\parallel[001]$, the ESR results suggest a 20\% reduction of the $g$-factor between $T=1.8$ and 3.2~K that can be clearly excluded based on our neutron-scattering data in Fig.~\ref{Fig:TempIndep}\,(b). This apparent contradiction is resolved by plotting the position of the ESR line for different temperatures and overlaying it on the phase diagram of CeB$_6$ for the same field direction \cite{NakamuraGoto95}, see Fig.~\ref{Fig:PhaseDiagESR}. Due to the technical restrictions of the ESR technique, the measurements have to be performed in a resonator with a fixed frequency (in this case, 60~GHz or 0.25~meV), and the microwave absorption is measured as a function of magnetic field. The red rectangle in Fig.~\ref{Fig:PhaseDiagESR} marks the region in the $B$-$T$ parameter space covered by the data in Refs.~\cite{SemenoGilmanov16, SemenoGilmanov17, Gilmanov19}, while the blue squares mark the positions of the 60~GHz resonance at different temperatures. In this representation, it becomes obvious that the field scans in the low-temperature region cross the phase boundary between the AFM and AFQ phases, where the magnetic excitation spectrum becomes quasielastic \cite{PortnichenkoDemishev16} and gets broadened by the critical fluctuations of both order parameters. In the proximity to a phase transition, the dependence of the resonance energy on magnetic field loses its linearity, which prohibits a reliable determination of the $g$-factor from a single measurement at one fixed frequency. We can therefore conclude that the temperature dependence of the ESR spectrum reported in Ref.~\cite{SemenoGilmanov16} results primarily from the proximity to phase~III, rather than from a change in the $g$-factor within phase~II, which we have shown to be temperature independent. Moreover, given that there are at least three magnetic resonances revealed in our INS measurements [Fig.~\ref{Fig:SummaryField}\,(d)], which cannot be clearly resolved at small fields, the ESR $g$-factor should correspond to an average slope of all three resonances. Taking the $g$-factor value of $\sim$\,1.4 at the maximal temperature of 3.2~K, which should be more reliable because the AFM phase is fully suppressed at this temperature, we find it to be in reasonable quantitative agreement with the average slope of the three INS resonances. We hope that these considerations will motivate additional ESR experiments in higher magnetic fields, which can be then directly compared with the presented INS data for $\mathbf{B}\parallel[001]$.

\vspace{-2pt}\section{Theory of multipolar excitations}\vspace{-3pt}
\label{Sec:Theory}

\subsection{Localized model for C\lowercase{e} $4f^1$ and multipolar HO}\vspace{-2pt}
\label{sec:model}

\begin{table*}[t!]
\begin{tabular}[c]{llll}
\toprule
degeneracy \;\;\;\;&$O_h$ rep., multipole\;\;\;\;\ \;\;\; & Stevens notation \;\;\;\;\;\;\;\;\;\;\;\;\;\;\; & pseudo-spin form  \\
\midrule	
3 &$\Gamma^-_4$, dipole & $J_x$  & $\frac{7}{6}(\sigma_x+\frac{2}{7}\eta_x)=\frac{7}{6}\sigma_x-\frac{1}{3}\tau_z\sigma_x+\frac{1}{\sqrt{3}}\tau_x\sigma_x$\\
&                   & $J_y$  & $\frac{7}{6}(\sigma_y+\frac{2}{7}\eta_y)=\frac{7}{6}\sigma_y-\frac{1}{3}\tau_z\sigma_y-\frac{1}{\sqrt{3}}\tau_x\sigma_y$ \\
&                   & $J_z$  & $\frac{7}{6}(\sigma_z+\frac{2}{7}\eta_z)=\frac{7}{6}\sigma_z+\frac{2}{3}\tau_z\sigma_z$  \\
3 &$\Gamma^+_5$, quadrupole & $O_{yz}=\frac{\sqrt{3}}{2}(J_yJ_z+J_zJ_y)$  & $\tau_y\sigma_x=\frac{1}{4}O_{yz}$\\
&                        & $O_{zx}=\frac{\sqrt{3}}{2}(J_zJ_x+J_xJ_z)$  & $\tau_y\sigma_y=\frac{1}{4}O_{yz}$\\
&                        & $O_{xy}=\frac{\sqrt{3}}{2}(J_xJ_y+J_yJ_x)$  & $\tau_y\sigma_z=\frac{1}{4}O_{yz}$\\
1 &$\Gamma^-_2$, octupole    & $T_{xyz}=\frac{\sqrt{15}}{6}\overline{J_xJ_yJ_z}$  & $\tau_y=\frac{\sqrt{5}}{45}T_{xyz}$\vspace{1pt}\\
\bottomrule
\end{tabular}
\caption{Table of 7 of the 15 multipoles of $\Gamma_8$ quartet in different representations: Stevens notation using total angular momentum components $J_\alpha$, $\alpha=x,y,z$, or pseudospin components $\sigma_\alpha$, $\tau_\alpha$. In the first three rows we used the multipole vector $\boldeta=(-\tau_z\sigma_x+\sqrt{3}\tau_x\sigma_x,~-\tau_z\sigma_y-\sqrt{3}\tau_x\sigma_y,~2\tau_z\sigma_z)$. The components of the total angular momentum are the linear combination $J_\alpha=\sum_n\lambda^n_\alpha X_n$ where coefficients $\lambda^n_\alpha$ can be read off in the last column. In the last row, symmetrization (summation over all permutations of $x$,\,$y$,\,$z$) is denoted by a bar.}
\label{tbl:multipole}
\end{table*}

We will now turn to the theoretical interpretation of the presented results and the discussion of the corresponding theoretical model that can explain the observed dependence of multipolar excitations on the magnitude and direction of the magnetic field. The localized 4$f$ model for the HO in \Ce\ has been proposed \cite{ohkawa:85} and investigated with respect to possible multipolar order \cite{shiina:97}. Its basic features will be briefly summarized here as a basis for interpreting experimental findings. It starts from the observation that the CEF splitting of $4f^1$ Ce$^{3+}$ $\Gamma_8$ ground state quartet and excited $\Gamma_7$ state is very large and therefore the latter may be neglected at low temperatures \cite{loewenhaupt:86, thalmeier:19, SundermannChen17}. In terms of free $4f^1 (J=\frac{5}{2})$ states $|M\ket\;(|M|\leq J) $ the quartet states $|\tau\sigma\ket$ $(\tau=\pm, \sigma=\ua\da)$ are given by
\begin{align}
\begin{split}
|+\ua\ket&=\sqrt{\textstyle\frac{5}{6}}|+{\textstyle\frac{5}{2}}\ket+\sqrt{\textstyle\frac{1}{6}}|-{\textstyle\frac{3}{2}}\ket; ~~~~~ |-\ua\ket=|{\textstyle+\frac{1}{2}}\ket,\\
|+\da\ket&=\sqrt{\textstyle\frac{5}{6}}|-{\textstyle\frac{5}{2}}\ket+\sqrt{\textstyle\frac{1}{6}}|+{\textstyle\frac{3}{2}}\ket; ~~~~~ |-\da\ket=|{\textstyle-\frac{1}{2}}\ket.
\end{split}
\label{eqn:CEF}
\end{align}
The fourfold degenerate ground-state manifold may be thought of being composed of two doublets of different $\Gamma_6$ and $\Gamma_7$ ``orbital'' symmetry, each being twofold Kramers degenerate. Therefore they can be presented by two types of pseudo-spins, namely $\boldtau$ for the two different orbitals and $\boldsigma$ for the Kramers degeneracy. The complete set of operators
\bea
\{X_n\}=\{\sigma_\alpha, \tau_\alpha, \sigma_\alpha\tau_\beta\}
\label{eqn:multipoleop}
\eea
with $n=1$--$15$ and $\alpha,\beta=x,y,z$ is constructed from the Pauli matrices of pseudo-spins. They constitute a basis set for the fifteen multipolar moments, acting in the space of $\Gamma_8$ quartet states, which are irreducible combinations of the $X_n$ in $O_h$ cubic symmetry. In terms of these variables the model Hamiltonian describing the effective intersite coupling of $\Gamma_8$ multipoles may be written as
\begin{multline}
\mathcal{H}=D\sum_{\langle ij\rangle}[(\boldtau_i\cdot\boldtau_j)+(\boldsigma_i\cdot\boldsigma_j)+4(\boldtau_i\cdot\boldtau_j)(\boldsigma_i\cdot\boldsigma_j)\,+\\
\epsilon_{\rm Q}4\tau_i^y\tau_j^y(\boldsigma_i\cdot\boldsigma_j)+\epsilon_{\rm O}\tau_i^y\tau_j^y]-\frac{7}{6}g\mu_{\rm B}\sum_i(\boldsigma_i+\frac{2}{7}\boldeta_i)\cdot{\bf H}.
\label{eqn:HAM}
\end{multline}
The first term is a SU(4) ``supersymmetric'' nearest-neighbor intersite interaction on the simple-cubic Ce-sublattice (sites $i,j$) of strength $D$, which has no bias for any of the fifteen multipoles as primary order parameter.
The second term describes the symmetry-breaking terms that favor quadrupolar or octupolar order depending on the size of the parameters $\boldepsilon=(\epsilon_{\rm Q},\epsilon_{\rm O})$, which correspond to $\Gamma^+_5$-type quadrupoles and $\Gamma^-_2$-type octupoles, respectively. The $\pm$ denotes even/odd behavior under time reversal. The last term is the Zeeman energy in pseudospin representation, where $\mathbf{H}=\mathbf{B}/\mu_0$ is the external magnetic field. The equivalence of multipole representation in terms of conventional total angular momentum operators $\bJ$ and the pseudospin expressions is presented in Table \ref{tbl:multipole}.

The previous conclusion from critical field anisotropy \cite{shiina:97}, field-induced neutron diffraction \cite{erkelens:87}, NMR results \cite{shiina:98}, and resonant x-ray scattering \cite{matsumura:09} indicate that HO may be well described as a primary AFQ $\Gamma^+_5$ order with a propagation vector $\bQ=(\pi,\pi,\pi)$ and a secondary strongly field-induced octupolar $\Gamma^-_2$ order parameter at the same wave vector. This may be achieved by choosing $\boldepsilon=(0.5,0.5)$ \cite{shiina:03} and should be considered as an appropriate set for \Ce~\cite{thalmeier:19}.

\vspace{-5pt}\subsection{RPA calculation of magnetic excitations in the multipolar-ordered phase}\vspace{-2pt}
\label{sec:RPA}

For the model in Eq.~(\ref{eqn:HAM}), the magnetic excitation spectrum of the multipolar-ordered phase as measured by INS may be calculated with complementary generalized Holstein-Primakoff approach \cite{shiina:03} or multipolar response function formalism in the RPA approach~\cite{thalmeier:98, thalmeier:03}; the latter will be used here. As the first step, one has to calculate the effective molecular fields $\langle X^n_A\rangle \pm \langle X^n_B\rangle$ (uniform and staggered) of each multipole basis operator where $s=A,B$ denote the simple-cubic sublattices of the antiferro-type HO defined by ordering vector $\bQ$. The CEF states are mixed by the molecular fields into new eigenstates $|\nu s\rangle_i$ with energies $E^s_\nu$ at every sublattice site $(s,i)$. In terms of standard basis operators a$^s_{\nu\mu}$(i)= $|\nu s\rangle_i\langle\mu s|_i$ $(\nu,\mu=1,4)$ for the molecular field states the mean field approximation to Eq.~(\ref{eqn:HAM}) is then given by
\begin{multline}
\mathcal{H}=\sum_{i,\nu,s}E^s_\nu a^s_{\nu\nu}(i)\,-\\
\frac{1}{2}\sum_{\bra ij\ket ss'}\sum_{\nu\nu',\mu\mu'}(\bM^s_{\nu\mu}\cdot\bD^{\phantom{s}}_{ss'}\cdot\bM^{s'}_{\nu'\mu'})\,a^s_{\nu\mu}(i)\,a^{s'}_{\nu'\mu'}(j),
\label{eqn:HAMMF}
\end{multline}
where $\bM$ is a 15-component vector of matrix elements for the multipole operators defined by $M^{ns}_{\nu\mu}$ = $\bra\!\nu,s|X^n_{is}|\mu,s\ket$ and the inter-sublattice multipole $15 \times 15$ diagonal interaction
matrix is $\bD_{AB}(\bq)=\bD_{BA}(\bq)=-2zD\gamma_\bq\boldLambda$. Here $\gamma_{\bq}$=$\frac{1}{3}(\cos q_x +\cos q_y +\cos q_z)$, $z=6$, and $\Lambda(n,n')=\Lambda(n)\delta_{n,n'}$ with $\Lambda(5) = 1+\epsilon_{\rm O}$
($\Gamma^-_2$ octupole), $\Lambda(\text{8--10}) = 1+\epsilon_{\rm Q}$ ($\Gamma^+_5$ quadrupole), and $\Lambda(n)=1$ otherwise (all other multipoles). Defining the thermal occupations of mean-field eigenstates by $n^s_\nu=Z_s^{-1}\exp(-E^s_\nu/T)$ with $Z_s=\sum_\mu\exp(-E^s_\mu/T)$, the bare $15 \times 15$ multipolar susceptibility for each sublattice $s=A,B$ may be written as\vspace{-2pt}
\bea
\chi^s_{0nl}(\omega)=\sum_{\nu\mu}\frac{M^{ns}_{\nu\mu}M^{ls}_{\mu\nu}}{E^s_\mu-E^s_\nu-\omega +{\rm i}\gamma_{\mu\nu}}(n^s_\nu-n^s_\mu).
\label{eqn:baresus}
\eea
The collective response of all 15 $\Gamma_8$ multipoles for the two sublattices is then described by the $30 \times 30$ RPA susceptibility matrix\vspace{-5pt}
\bea
\boldchi(\bq,\omega)=[\bun -\boldchi_0(\omega)\bD({\bq})]^{-1}\boldchi_0(\omega).
\label{eqn:RPASUS}
\eea
From its elements we construct the physical-moment (\bJ) cartesian $3 \times 3$ susceptibility matrix according to
\bea
\chi_{\alpha\beta}(\bq,\omega)=\sum_{\!\!\!ss'\!,\,nm\!\!\!}\lambda^n_\alpha\lambda^m_\beta\chi_{nm}^{ss'}(\bq,\omega),
\eea
where $\lambda^n_\alpha$ are the coefficients of $J_\alpha$ in the pseudo-spin representation (Table \ref{tbl:multipole}). Then the dynamical {\it dipolar} structure function seen in INS may be written as\vspace{-3pt}
\begin{multline}
S(\bQ,\omega;\mathbf{B})=\frac{1}{\pi}[1-{\rm e}^{\hbar\omega/kT}]\\
\times\sum_{\alpha\beta}[\delta_{\alpha\beta}-\hat{\bQ}_\alpha\hat{\bQ}_\beta]\,{\rm Im}\chi_{\alpha\beta}(\bq,\omega;\mathbf{B}),\vspace{-2pt}
\label{eqn:strucfac}
\end{multline}
which depends parametrically on field strength and direction. Here $\bQ = \bq+\bK$ is the total momentum transfer in INS with \bK\ denoting a reciprocal lattice vector and $\hat{\bQ} = \bQ/|\bQ|$. The prefactor under the sum projects to the contributions with momentum transfer perpendicular to the dipolar moment \bJ. In the following the structure function will be computed for various model parameters, field strengths and continuous rotations of field direction and compared to the experimental INS scattering results.

\vspace{-3pt}\subsection{Dispersion of multipolar modes in phase~II}\vspace{-3pt}
\label{Sec:TheoryDispersion}

First we discuss the dispersion of multipolar modes for a fixed magnetic field $\mathbf{B}$. In the model calculations, we only include the AFQ order of phase~II, the effect of the complicated dipolar magnetic order (phase III) \cite{erkelens:87, thalmeier:19} at a different wave vector is not treated here. An attempt to this purpose has been made in \cite{kusunose:01}. The model parameters are given by $\boldepsilon = (0.5,0.5)$ \cite{shiina:03, thalmeier:03}, corresponding to the composite degenerate $\Gamma^+_5$ quadrupole $(O_{yz},O_{zx},O_{xy})$ and $\Gamma^-_2$ $(T_{xyz})$ octupole order. It should be noted that the precise values of $\epsilon_{\rm Q},\epsilon_{\rm O}$ are not known. The accidental degeneracy $\epsilon_{\rm Q}=\epsilon_{\rm O}$ is assumed to minimize the number of parameters. This model is known to reproduce the magnetic-field\,--\,temperature phase diagram of \Ce\ and its field anisotropies well, including the fluctuation effects beyond the mean-field approximation~\cite{shiina:02} as well as the essential NMR results~\cite{shiina:98}. In particular, $\epsilon_{\rm Q}=\epsilon_{\rm O}$ can explain the strongly convex shape of induced octupole $T_{xyz}(\mathbf{B})$ field dependence observed directly in RXS experiments~\cite{matsumura:12}.

Furthermore $k_{\rm B}T_0=2zD=0.41$~meV or $T_0=4.74$~K defines the interaction energy and temperature scale. Following the notation of Ref.~\cite{thalmeier:03}, we can also introduce the dimensionless field strength $h'=h/T_0$ with $h=(7/6)g_J\mu_{\rm B}H/T_0$ (assuming $k_{\rm B}\equiv 1$), which can be expressed as $h'=0.672B[\text{T}]/T_0[\text{K}]$ using physical units for $B$~and~$T_0$. Note that the mean-field transition temperature of the model is $T^{\rm mf}_{\rm Q}(h'\!=\!0)=(1\!+\!\epsilon_{\rm Q})T_0 = 7.12$~K, which is considerably larger than the experimental transition temperature $T^{\rm mf}_\text{exp}=3.2$~K, because the latter is strongly suppressed by fluctuations \cite{shiina:02}.

\begin{figure}[t!]
\includegraphics[width=1.01\columnwidth]{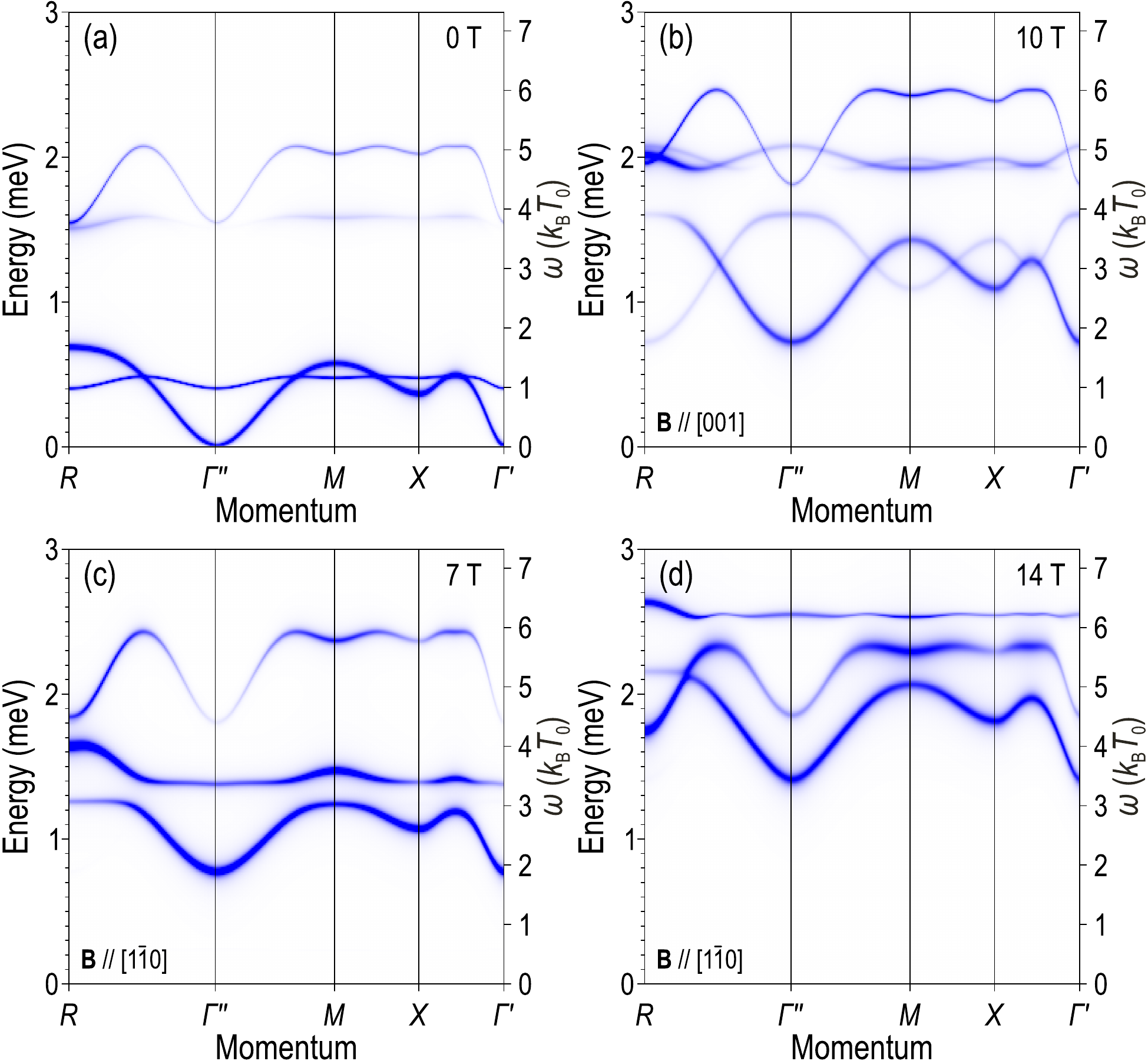}
\caption{The calculated dipolar structure function of magnetic excitations along the continuous polygonal path in momentum space: $R(\frac{1}{2}\frac{1}{2}\frac{1}{2})$\,--\,$\Gamma''(110)$\,--\,$M(\frac{1}{2}\frac{1}{2}0)$\,--\,$X(\frac{1}{2}00)$\,--\,$\Gamma'(100)$. (a)~For zero magnetic field; (b)~for $h'=1.45$ ($B\approx10$~T) applied along the [001] direction; (c,\,d)~for fields $h'=1$ ($B\approx7$~T) and $h'=2$ ($B\approx14$~T) applied along the $[1\overline{1}0]$ axis. The left part (segment $R$\,--\,$\Gamma''$\,--\,$M$) of panel (d) and the middle part (segment $\Gamma''$\,--\,$M$) of panel (b) are therefore approximately equivalent to the experimental Figs.~\ref{Fig:DispersionField}\,(a,\,b), respectively.\vspace{-4pt}}
\label{Fig:TheoryDisp}
\end{figure}

The calculated dynamical RPA dipolar structure function $S(\mathbf{Q},\omega)$ for \Ce\ is shown in Fig.~\ref{Fig:TheoryDisp}. The polygonal path in momentum space and the strength and magnitude of the magnetic field used in the calculation were chosen to enable comparison with the experimental data. The dispersion of multipolar excitations is in excellent agreement with the results of the alternative HP approach, including the intensities of modes as a function of $\mathbf{Q}$ \cite{shiina:03, thalmeier:03}. Because there are two sublattices and locally three excited CEF states, in the AFQ/AFO-type ordered phase six excitation branches will appear prominently at low temperature that are mostly visible, e.g. in Fig.~\ref{Fig:TheoryDisp}\,(b--d) (for $h=0$ there are additional degeneracies). Roughly the six branches can be arranged in two groups: The low-energy branches with a strong field dependence are stabilized directly by the Zeeman term, whereas the remaining high-energy branches are mainly stabilized by the strongly field-induced octupolar molecular field \cite{shiina:03}.

When temperature approaches $T^{\rm mf}_{\rm Q}(h')$ from below at a constant field, the high-energy modes are expected to collapse due to the reduced octupolar order, while the low-energy modes should be less affected. Complementary, when temperature is kept constant much below $T^{\rm mf}_{\rm Q}$ and the field is reduced to zero, the high-energy modes change little, while the low-energy modes are shifted downwards. Due to the threefold degenerate $\Gamma^+_5$ order, a Goldstone mode then appears for $h'=0$ at the $\Gamma$ point, see Fig.~\ref{Fig:TheoryDisp}\,(a). At finite field, the degeneracy is lifted by choosing a linear combination of the three $\Gamma_5^+$ quadrupoles as the ground state (see Sec.~\ref{sec:fieldangular}), leading to a $\Gamma$-point anisotropy gap in the excitation spectrum in Figs.~\ref{Fig:TheoryDisp}\,(b--d). It should be mentioned that the relative field independence of the higher-energy modes is a consequence of the accidental degeneracy $\epsilon_{\rm Q}=\epsilon_{\rm O}$ assumed in the model \cite{thalmeier:03} which is consistent with the RXS results~\cite{matsumura:12}.

The color plots in Fig.~\ref{Fig:TheoryDisp} show not only the dispersion, but also the expected intensity of the multipolar excitations that can be directly compared with the INS experiments presented in Sec.~\ref{SubSec:Dispersion}. The INS data in Fig.~\ref{Fig:DispersionField}\,(a), measured along the $R$\,--\,$\Gamma''$\,--\,$M$ path for $\mathbf{B}\parallel[1\overline{1}0]$, should be directly compared with the left half of Fig.~\ref{Fig:TheoryDisp}\,(d), whereas the experimental color map in Fig.~\ref{Fig:DispersionField}\,(a), measured along the $\Gamma''$\,--\,$M$ direction for $\mathbf{B}\parallel[001]$, should be compared with the corresponding section of Fig.~\ref{Fig:TheoryDisp}\,(b). As mentioned in the introduction, we expect no quantitative agreement with experiment away from the $\Gamma$ point, as our theoretical model neglects important interactions beyond the nearest neighbors, yet even such an oversimplified approach reveals some qualitative similarities with the measured data. For instance, the crossing of the two low-energy branches along the $\Gamma''$\,--\,$M$ direction for $\mathbf{B}\parallel[001]$ and along the $R$\,--\,$\Gamma''$ direction for $\mathbf{B}\parallel[1\overline{1}0]$ are quite well captured by the model.

Note that under an external field, the intensity distribution no longer follows the cubic symmetry of the lattice. The dynamic structure factor depends on the relative orientation of the $\mathbf{Q}$ and $\mathbf{B}$ vectors, as illustrated in Figs.~\ref{Fig:TheoryB100vsB001} and \ref{Fig:TheoryB1m10vsB110} in the Supplemental Material \cite{SupplementalMaterial}. For instance, the intensity of one of the high-energy modes at $\mathbf{Q}=\Gamma'(100)$ fully vanishes if the magnetic field is applied parallel to the $\mathbf{Q}$ vector, $\mathbf{B}\parallel[100]$, but is finite for equal but orthogonal field \mbox{$\mathbf{B}\parallel[001]\perp\mathbf{Q}$} (Fig.~\ref{Fig:TheoryB100vsB001}). Similar changes in intensity are seen near $\mathbf{Q}=\Gamma''(110)$ for magnetic fields along $[1\overline{1}0]\perp\mathbf{Q}$ or $[110]\parallel\mathbf{Q}$ (Fig.~\ref{Fig:TheoryB1m10vsB110}).

\vspace{-2pt}\subsection{Dependence of multipolar excitations on the field strength and direction}\vspace{-2pt}
\label{sec:field}

\begin{figure}[t]
\includegraphics[width=\columnwidth]{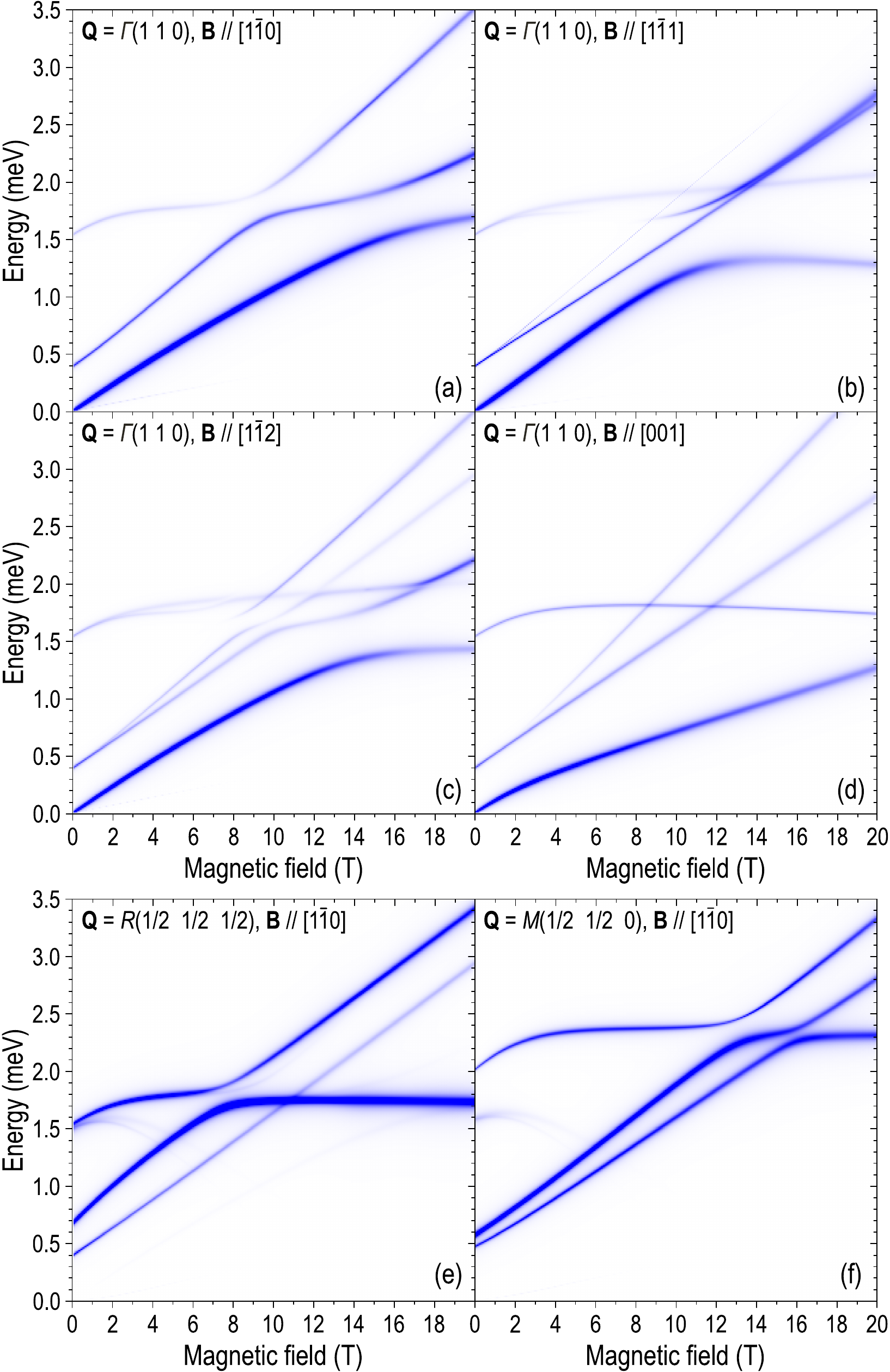}
\caption{Calculated magnetic-field dependence of multipolar excitations and their intensities as expected in an INS measurement (dipolar structure function). (a--d)~At the $\Gamma''(110)$ point under different field directions: $[1\overline{1}0]$, $[1\overline{1}1]$, $[1\overline{1}2]$, and $[001]$, respectively. (e,\,f)~At the $R(\frac{1}{2}\frac{1}{2}\frac{1}{2})$ and $M(\frac{1}{2}\frac{1}{2}0)$ points for $\mathbf{B}\parallel[1\overline{1}0]$. These configurations are identical to those in Fig.~\ref{Fig:SummaryField}.}
\label{Fig:TheoryFieldDep}
\end{figure}

We will now consider the magnetic-field dependence of mode energies for a fixed momentum transfer $\mathbf{Q}$. To minimize the number of free parameters, we will restrict our consideration to the high-symmetry points of the BZ that were investigated experimentally. As pointed out in the introduction, the benefit of this approach is that for a fixed $\mathbf{Q}$ corresponding to a symmetry point, the various interactions enter only in specific sums and not their individual values which should be much less sensitive to the unavoidable inaccuracies in the description of the multipolar interaction Hamiltonian. For a fixed $\mathbf{Q}$, the momentum-dependent part of the model can result in an energy offset of the field dependence, which we expect to vanish at the zone center ($\Gamma$ point) at least for the low-energy excitation branches. Looking at the multipolar excitations in field space (magnitude and direction of $\mathbf{B}$) for a fixed momentum is a change of viewpoint that represents a complementary approach as compared to the previous theoretical \cite{shiina:03, thalmeier:03} and experimental \cite{jang:14, PortnichenkoDemishev16} investigations.

\vspace{-3pt}\subsubsection{Variation of mode energies with field strength}\vspace{-3pt}
\label{sec:fieldstrength}

First, we consider the continuous dependence of mode energies at the high-symmetry points of the BZ on field strength, which is illustrated in Fig.~\ref{Fig:TheoryFieldDep}. The panels in this figure are arranged in analogy to the experimental Fig.~\ref{Fig:SummaryField} and were calculated for the same $\mathbf{Q}$ vectors and field configurations. The four upper panels (a--d) show the multipolar excitation energies at the $\Gamma''(110)$ point for field directions $[1\overline{1}0]$, $[1\overline{1}1]$, $[1\overline{1}2]$, and $[001]$. This corresponds to a rotation of the field direction in the plane orthogonal to the $\mathbf{Q}$ vector. In addition, a similar field dependence for $\mathbf{B}\parallel\mathbf{Q}$ (which was not investigated experimentally) is presented in Fig.~\ref{Fig:TheoryB110}\,(d) of the Supplemental Material \cite{SupplementalMaterial}.

\begin{figure}[b]
\includegraphics[width=1.0\columnwidth]{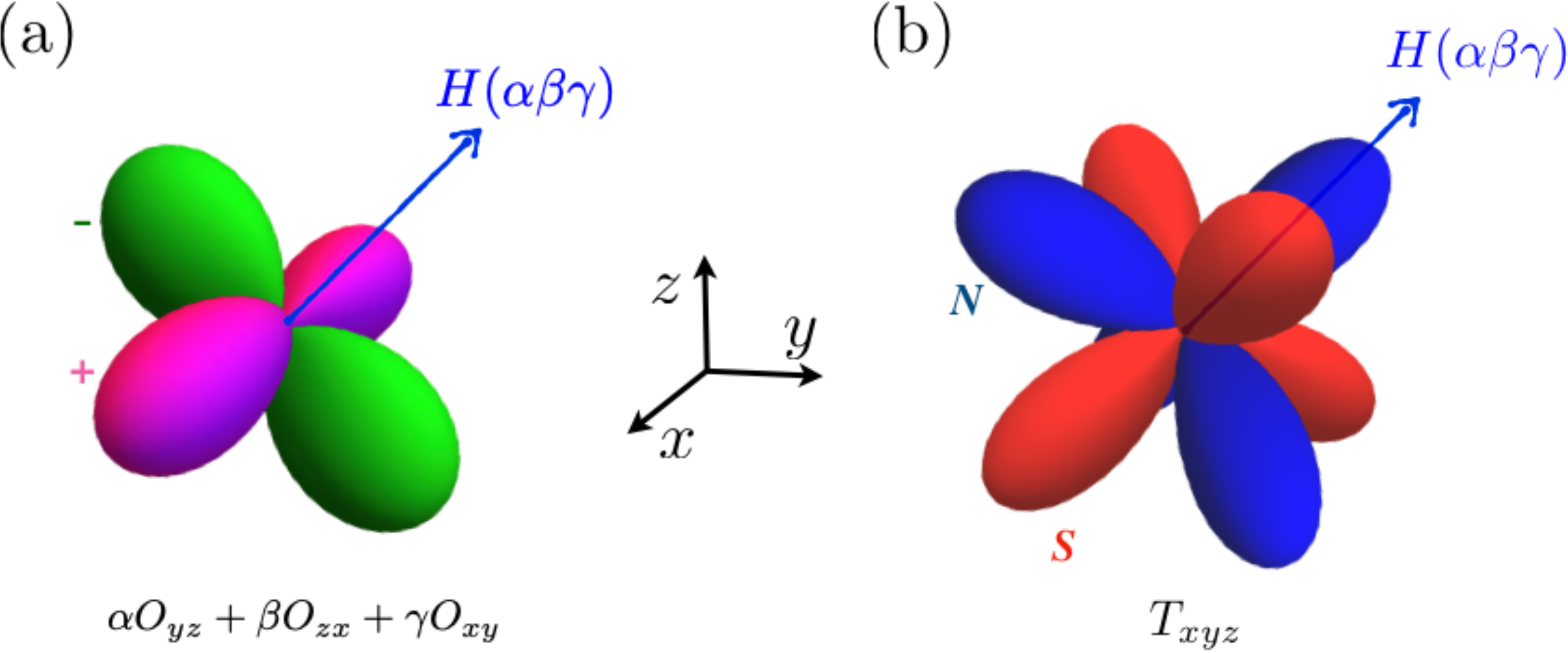}
\caption{Sketch of the quadrupole (left) and octupole (right) order parameter in a rotating field that points along the unit vector $(\alpha\beta\gamma)$ obtained from the real-space equivalent tesseral harmonics of the Stevens operators in Table \ref{tbl:multipole}. The distance from center to the lobes is proportional to the charge/spin-density of the even $\Gamma_5^+$ quadrupole or odd $\Gamma_2^-$ octupole. The total charge or spin moment is zero. Therefore equal regions with $\pm$ charge or (N/S) spin densities (different colors for each) appear. When the field unit vector $(\alpha\beta\gamma)$ rotates, the quadrupole charge density rotates with it such that it remains in the plane perpendicular to $\mathbf{H}$. This is due to the 3-fold degeneracy of $\Gamma_5^+$. In contrast, the non-degenerate $\Gamma_2^-$ octupole spin density remains fixed with respect to the crystal axes when $\mathbf{H}$ rotates. As a function of field strength $H$, the primary quadrupole order parameter is little affected, while the induced octupole shrinks to zero for $H\rightarrow 0$.\vspace{-4pt}}
\label{Fig:QuadruOctu}
\end{figure}

There are very pronounced differences in the mode energies for different field directions, related to the behavior of the $O_{xy}$ quadrupoles and the $\Gamma_2^-$ octupoles, that form primary and secondary (field-induced) order parameters of phase~II, with an applied field. The $O_{xy}$ and equivalent $O_{yz}$ and $O_{zx}$ quadrupoles are selected from the $\Gamma^+_5$ manifold as the primary AFQ order parameter. Their threefold degeneracy (Table~\ref{tbl:multipole}) enables a free rotation of the charge density of the quadrupole such that its lobes always point in the directions orthogonal to the field, as shown in Fig.~\ref{Fig:QuadruOctu}. For a magnetic field pointing along the direction defined by the unit vector $\mathbf{H}/|\mathbf{H}|=(\alpha\beta\gamma)$, the primary order parameter can be written as the linear combination $\alpha O_{yz}+\beta O_{zx}+\gamma O_{xy}$~\cite{thalmeier:19}. In contrast, the field-induced octupole $T_{xyz}$ is nondegenerate and therefore fixed with respect to the crystal axes when $\mathbf{H}$ rotates.

As mentioned before, the high-energy modes $\omega\simeq 4T_0\approx1.6$~meV are stabilized by the octupolar molecular field and are hardly influenced by the applied magnetic field. They can be seen as horizontal, approximately field-independent lines in Fig.~\ref{Fig:TheoryFieldDep}. On the other hand, the low-energy modes are stabilized directly by the Zeeman term and show an approximately linear increase in frequency and splitting with the field, with the slope given by the effective $g$-factor that can be anisotropic with respect to the field direction \cite{SemenoGilmanov16, SemenoGilmanov17, Gilmanov19}. At sufficiently high fields, these modes intersect with the high-energy modes, which leads to a complex hybridization pattern with multiple avoided mode crossings. This can lead to drastic changes in the slope and a redistribution of intensity between different modes above a certain field threshold, possibly leading to the appearance of an additional ESR line \cite{DemishevSemeno08} and to the second higher-energy peak in the INS spectrum described in Sec.~\ref{Sec:INS110}.

To simulate how the specific choice of the $\mathbf{Q}$ vector at which the $\Gamma$-point spectra are measured in an INS experiment influences the relative intensity of the multipolar modes, in Figs.~\ref{Fig:TheoryB001} and \ref{Fig:TheoryB110} in the Supplemental Material \cite{SupplementalMaterial} we compare the calculated field dependencies for $\mathbf{B}\parallel[001]$ at $\mathbf{Q}=(100)$, $(001)$, $(110)$, and $(111)$, and for $\mathbf{B}\parallel[1\overline{1}0]$ at $\mathbf{Q}=(001)$, $(110)$, $(1\overline{1}0)$, and $(111)$. The observed changes in the dynamical structure factor are substantial, which may lead in the extreme case to the complete disappearance of some of the modes from the spectrum, e.g. for $\mathbf{B}\parallel\mathbf{Q}=(001)$ in Fig.~\ref{Fig:TheoryB001}. This suggests a possible way of identifying the theoretically predicted excitations in the experimental spectrum by the behavior of their structure factor measured for several equivalent $\mathbf{Q}$ points.

Figures \ref{Fig:TheoryFieldDep}\,(e,\,f) also show similar field dependences for two other high-symmetry points, $R(\frac{1}{2}\frac{1}{2}\frac{1}{2})$ and $M(\frac{1}{2}\frac{1}{2}0)$, that were investigated experimentally. In spite of the limited INS data for these points, it is clear that the linear field dependence at the $R$ point in the experimental data (Fig.~\ref{Fig:SummaryField}) extends to much higher fields and does not exhibit the pronounced flattening of the strongest excitation near 8~T, suggested in the calculations. We attribute this discrepancy to the limitations of our model, which is not expected to be quantitatively accurate at wave vectors other than the zone~center.

\begin{figure}[t]
\begin{center}
\includegraphics[width=\columnwidth]{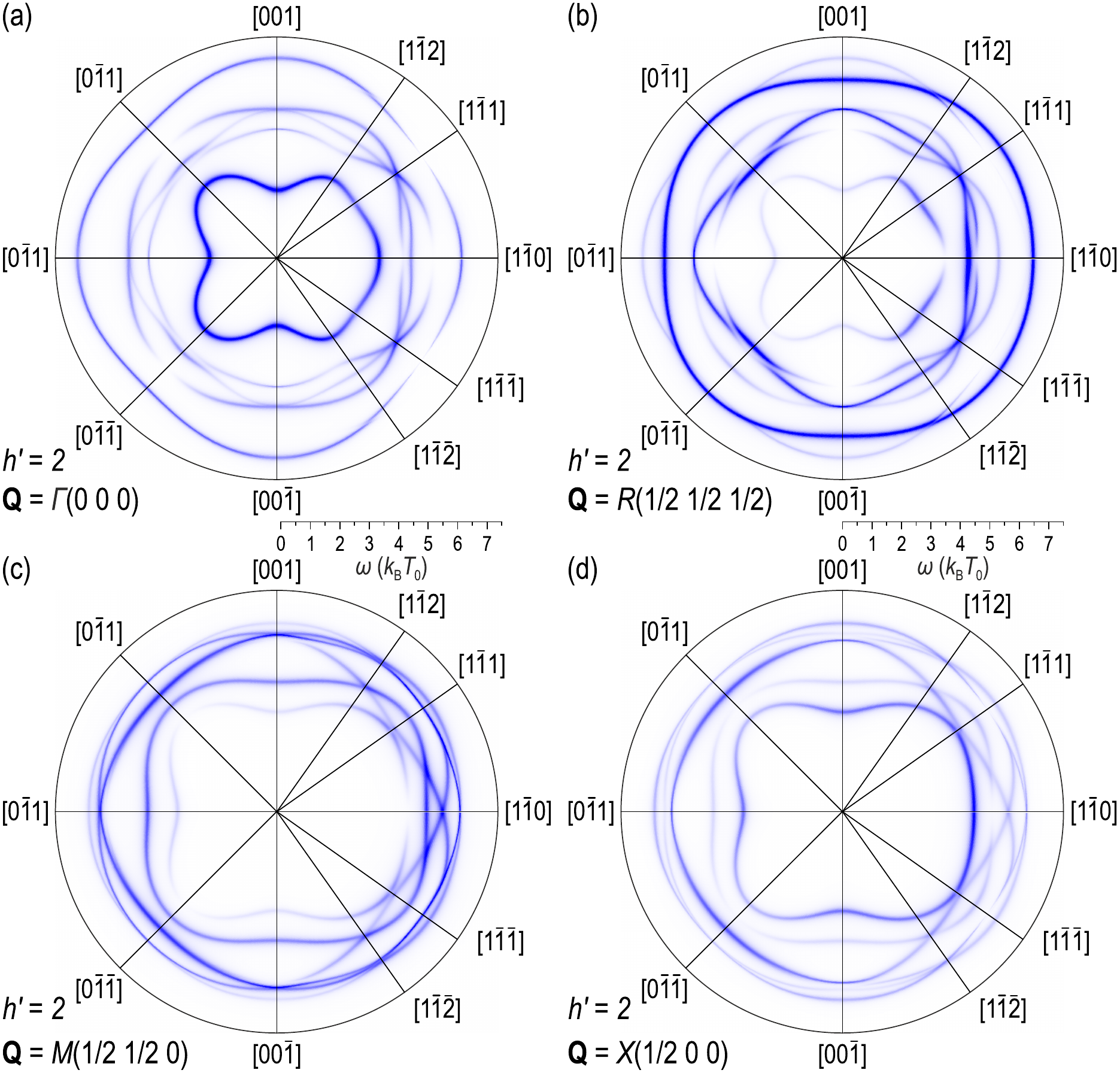}
\end{center}
\caption{Field-angular anisotropies of multipolar excitation branch\-es at the high-symmetry points of the cubic BZ, as indicated beside every panel. We use a polar representation, where the angle gives the field direction and the radius represents the mode frequency plotted from zero (center) to $7.5 T_0$ (outer circular boundary). For all plots, the magnetic field strength is fixed to $h'=2$ ($B\approx14$~T). Note that in this figure the field rotation is not in a single plane for the whole range of polar angles: The left and right halves of every plot correspond to the field rotation in the planes orthogonal to $[100]$ and $[110]$, respectively.}
\label{Fig:TheoryPolar}
\end{figure}

\vspace{-2pt}\subsubsection{Multipolar excitations under rotation of the field direction}\vspace{-3pt}
\label{sec:fieldangular}

Another result of our new approach to the multipolar excitations in \Ce\ concerns the dependence on continuous field rotation and is presented in the polar graphs of Fig.~\ref{Fig:TheoryPolar} for $h'=2$ ($B \approx 14$~T). In these figures, the radial coordinate represents the frequency of the excitation modes in the dipolar INS structure function of Eq.~(\ref{eqn:strucfac}). The angular coordinate defines the field-rotation angle between various cubic axes as indicated at the outer boundary of the polar plots. The rotation is continuous and closed, covering two planes orthogonal to $[100]$ and $[110]$, as shown in the left and right halves of every plot, respectively.

For each field direction given by the unit vector $(\alpha\beta\gamma)$, the AFQ order parameter $\alpha O_{yz}+\beta O_{zx}+ \gamma O_{xy}$ is selected from the $\Gamma_5^+$ manifold \cite{thalmeier:19} (see Fig.~\ref{Fig:QuadruOctu}), which is automatically included in the mean-field ground state of the model given by Eq.~(\ref{eqn:HAM}). The resulting field-angular mode anisotropies show various signatures that are worthwhile to mention and to look for in future experiments. First of all, the anisotropy of the low-energy modes that exhibit approximately linear field dependence at small fields represents the effective $g$-factor anisotropy that can be directly measured both in ESR (for the $\Gamma$ point) \cite{SemenoGilmanov16, SemenoGilmanov17, Gilmanov19} and in INS (for any $\mathbf{Q}$ vector). The high-energy mode frequency (large polar radii) is less sensitive to the magnetic field and therefore changes (expands) slower with increasing field strength than the low-energy modes (small polar radii). As a result, an increase in field strength leads to a hybridization of the low- and high-energy modes, which generally increases the complexity and anisotropic appearance of the field-angular variation of mode frequencies. We also note that the anisotropy patterns at the $\Gamma$ and $R$ points are quite similar. This is expected since both points have full cubic symmetry. Interestingly, the relative intensity of the low-energy and high-energy modes is interchanged when going from $\Gamma$ to $R$ and vice versa, in agreement with the experimental result in Fig.~\ref{Fig:DispersionField}\,(a). At the $X$ and $M$ points, the low-energy mode frequencies are larger (inner radii increased), but again we observe some intensity exchange, now among the low-energy modes themselves.

Altogether these polar field-angle anisotropy plots of multipolar excitations in \Ce\ at high-symmetry points represent a large set of information on mode positions and intensities that may be very useful for comparing with future experimental results and should give guidance where to look for the modes with largest intensity. If the model so far accepted for \Ce\ with $\boldepsilon=(0.5,0.5)$ is reasonable, some features of the field-anisotropy plots described above should be identified in INS experiment with quasi-continuous field rotation. Such a comparison becomes possible using the data presented in Sec.~\ref{Sec:ExpAnisotropy}.\vspace{-5pt}

\vspace{-3pt}\subsection{Comparison to the experimental results}\vspace{-3pt}
\label{sec:strength_exp}

\begin{figure}[b]
\includegraphics[width=\columnwidth]{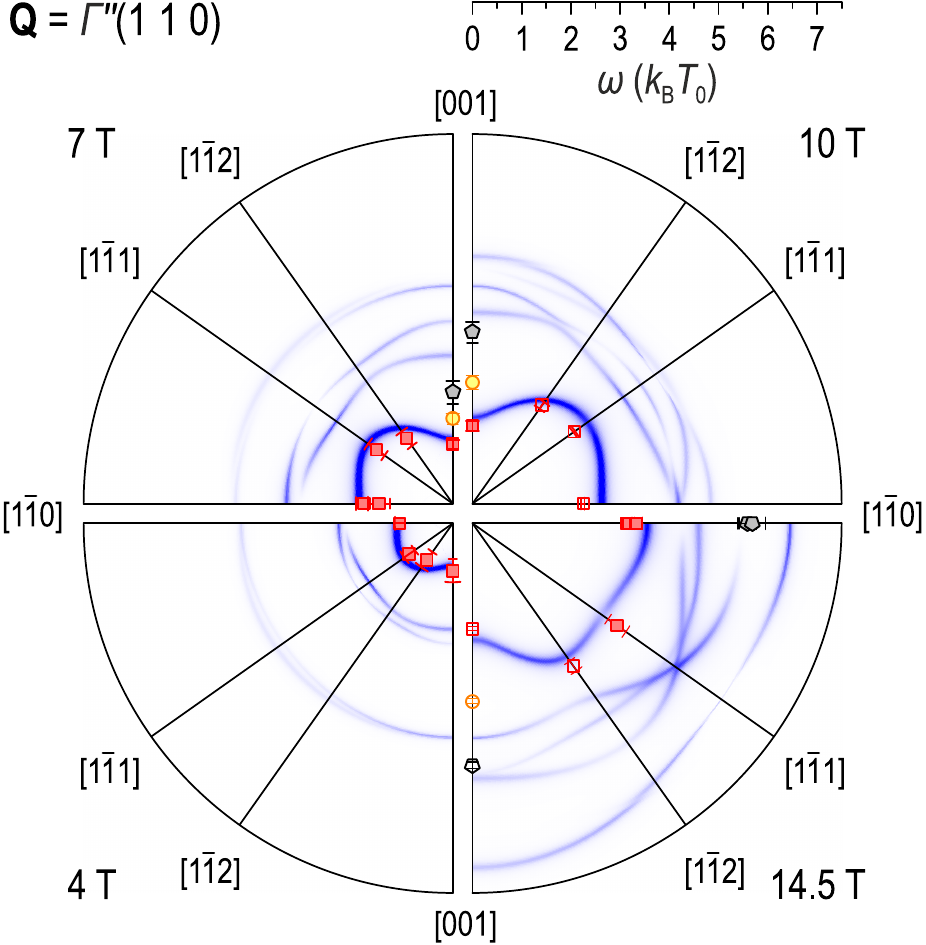}
\caption{Multipolar excitation branches for the $\Gamma''(110)$ point in polar representation, where the radial coordinate corresponds to the excitation energy, plotted from 0 to $7.5 T_0$, and the angular coordinate to the magnetic field direction, continuously rotating from $[001]$ to $[1\overline{1}0]$ in the plane orthogonal to $[110]$. The four segments of the figure show results for 4, 7, 10, and 14.5\,T. The color map shows the results of the calculation, with the experimental data points along high-symmetry field directions overlayed for comparison. The symbol shapes for different modes are consistent with those in Fig.~\ref{Fig:SummaryField}. Filled symbols correspond to direct measurements of peak position, whereas open symbols are the results of interpolation from the field dependences plotted in Fig.~\ref{Fig:SummaryField}.\vspace{-4pt}}
\label{Fig:PolarPlotComparison}
\end{figure}

A reliable comparison with experiment is so far only possible at the $\Gamma$ point, where detailed field-angular dependences of the INS spectra were obtained (see, for instance, Figs.~\ref{Fig:GammaRawScan110}, \ref{Fig:ThalesFLEXXbooster110}, \ref{Fig:SummaryField}, \ref{Fig:ComparisonFields}, \ref{Fig:TempIndep}). In Fig.~\ref{Fig:PolarPlotComparison}, the corresponding experimental peak positions for the fields of 4, 7, 10, and 14.5\,T are plotted on top of the calculated field-angular polar maps for the same magnetic fields. Every segment of the plot covers the same irreducible 90$^\circ$ range of field directions between $[001]$ and $[1\overline{1}0]$ in the plane orthogonal to $[110]$. The experimental data points were obtained either by directly fitting the peak positions from measurements at the corresponding field values (filled symbols) or via linear interpolation of the field dependences (open symbols). Different symbol shapes for different modes are consistent with those in Fig.~\ref{Fig:SummaryField}.

We see a remarkably good agreement of the calculation with the field-directional anisotropy of the most intense low-energy mode (resonance~A) with a quasi-linear field dependence, which has been followed experimentally over a large field range for all four high-symmetry directions of the magnetic field: $\mathbf{B}\parallel[001]$, $[1\overline{1}2]$, $[1\overline{1}1]$, and $[1\overline{1}0]$. At high magnetic fields of the order of 10~T, we observe a very considerable anisotropy of $\sim$\,60\% in the effective $g$-factor between its minimal and maximal values reached for the $[001]$ and $[1\overline{1}1]$ field directions, respectively. For lower fields, however, the anisotropy is reduced, which is a direct consequence of the initially nonlinear field dependence of the low-energy mode that can be seen in Fig.~\ref{Fig:TheoryFieldDep}\,(d). Indeed, the $g$-factor anisotropy reported from previous ESR measurements \cite{SemenoGilmanov16, SemenoGilmanov17, Gilmanov19}, which were performed at magnetic fields of $\sim$\,3~T, is several times smaller and constitutes a relative change of less than 10\% between the [100] and $[111]$ field directions. However, the overall shape of the anisotropic $g$-factor dependence turns out to be the same in ESR and INS measurements, if one considers only those ESR data that were measured at an elevated temperature away from the transition line to the AFM phase (see Fig.~\ref{Fig:PhaseDiagESR} and the discussion at the end of Sec.~\ref{Sec:ExpAnisotropy}). The same anisotropic field-angular dependence could be well described by a model proposed in \mbox{Refs.~\cite{Schlottmann12, Schlottmann13, Schlottmann18}} by Schlottmann.

Another experimental confirmation of the possibly nonlinear field dependence of the low-energy mode in Fig.~\ref{Fig:TheoryFieldDep}\,(d) for the field direction $\mathbf{B}\parallel[001]$ follows from the experimental field dependence plotted in Fig.~\ref{Fig:SummaryField}\,(d). This is the only field direction, for which an extrapolation of the high-field part of the field dependence (measured in phase~II, far from the AFM phase) for the position of the most intense low-energy peak results in a small but significant offset on the energy axis, in agreement with the theoretical model. For other field directions, such as $\mathbf{B}\parallel[1\overline{1}0]$ in Fig.~\ref{Fig:TheoryFieldDep}\,(a), the deviations from a linear dependence first start at much higher fields due to the hybridization with high-energy modes.

Figure \ref{Fig:PolarPlotComparison} also shows the positions of the new resonance peaks that have been so far measured only for $\mathbf{B}\parallel[1\overline{1}0]$ and $\mathbf{B}\parallel[001]$ in high magnetic fields. Their energies fall in the range where the theoretical model predicts multiple modes resulting from the hybridization of high- and low-energy excitations. With the available data, it is not possible to assign one of these modes uniquely to the experimentally observed resonances. However, the field dependence in Fig.\,\ref{Fig:TheoryFieldDep}\,(a) suggests that the high-energy mode (resonance~C) may correspond to the upper branch that gains intensity beyond the avoided mode crossing for $B\gtrsim10$~T and $\hslash\omega\gtrsim2$~meV. This field and energy range approximately corresponds to the region where resonance~C has been seen in our experiment. The predicted energy splitting between the two modes of $\sim$\,1~meV also roughly matches with the energy difference between the INS resonances A and C.

On the other hand, the model predicts no additional resonances at the $\Gamma$ point below the energy of resonance~A, which means that it is unable to explain the appearance of the resonance~B in ESR data at fields above 12~T \cite{DemishevSemeno08}. Theoretical considerations about the possible origin of this high-field resonance were proposed earlier by Schlottmann~\cite{Schlottmann12, Schlottmann13, Schlottmann18}, but the reason why it cannot be seen in INS measurements remains unclear. The dispersion of resonance~B, as well as its relationship to the multipolar excitations described in our work, also remain to be clarified.

\vspace{-2pt}\section{Summary and Discussion}\vspace{-2pt}
\label{sec:summary}

In summary, we have proposed an alternative approach to study magnetic excitations in a correlated-electron system characterized by an interplay of charge and orbital degrees of freedom, based on the measurements of INS spectra at fixed points in momentum space as a function of magnetic field and its direction. This approach is complementary to the conventional analysis of INS data, which primarily considers the dispersion of magnetic excitations as a function of momentum for a single magnetic-field direction. Measurements of the field-directional anisotropy are a well-established technique in many physical probes, such as thermodynamic or transport measurements, and it is therefore natural to extend this approach to the local spectroscopic probes, such as neutron spectroscopy. In complex systems characterized by strong spin-orbit coupling and exhibiting multipolar order parameters, INS measurements of the field-angular anisotropy of magnetic excitations can help in disentangling the ``local'' physics related to the field-dependent molecular field and the Zeeman splitting of the single-ion eigenstates from the effects of magnetic interactions that lead to the formation of dispersive collective modes with a pronounced momentum dependence.

We have applied the proposed method to the analysis of INS data on the long-studied material CeB$_6$, which exhibits complex low-temperature ordering phenomena and rich magnetic excitation spectra that have so far avoided a complete theoretical description in spite of many years of experimental and theoretical effort. A comparison of our experimental results presented in Sec.~\ref{Sec:Experiment} with the model calculations described in Sec.~\ref{Sec:Theory} reveals reasonable agreement of the excitation spectra in the vicinity of the zone center ($\Gamma$ point). In particular, the available model can reproduce the experimental $g$-factor anisotropy and can qualitatively explain the appearance of the new high-energy mode in the excitation spectrum that can only be observed at high magnetic fields. On the other hand, the model is much less accurate in describing the dispersion of multipolar excitations away from the zone center, as it neglects significant long-range RKKY couplings between the next-nearest and next-next-nearest neighbors. Potentially, combining INS data in momentum and field space may provide sufficient information to fit multiple interaction terms in a more realistic Hamiltonian of CeB$_6$ in a similar way, as it is routinely done for conventional dipolar magnets.

The vector field effectively adds three new dimensions to the parameter space of the INS measurement, making it exceedingly difficult to perform the full mapping of the parameter space, as it would require an unreasonably large amount of neutron beam time. This is perhaps one of the reasons why such an approach has not been applied to any other material until now, to the best of our knowledge. Another reason is more technical, related to the development of stable and reliable mechanical rotators that would be compatible with high-field magnets used in INS experiments~\cite{ThalerNorthen16}. In future, with the advent of new high-flux neutron sources and the development of suitable sample environment for the stable mechanical rotation of single crystals around a specific scattering vector, measurements similar to those presented here for CeB$_6$ can be accomplished much more efficiently and for a broad range of materials. The subsequent comparison of such measurements with the corresponding theoretical model of multipolar excitations is able to provide the most direct information about inter-multipolar interactions that stabilize magnetically hidden order and to probe the HO symmetry across the magnetic phase diagram. We hope that our present work will guide future developments in this direction.\vspace{-3pt}

\section*{Acknowledgments}

We thank Philippe Boutrouille (LLB), Ralf Feyerherm, Bastian Klemke, and Klaus Kiefer (HZB) for technical support during the experiments. A.\,A. and P.\,T. thank Ryousuke Shiina for help with the computations. Reduction of the TOF data was done using the \emph{Horace} software package \cite{ewings:16}. This project was funded by the German Research Foundation (DFG) under the individual grant No.~IN\,\mbox{209/3-2}, from the project C03 of the Collaborative Research Center SFB\,1143 at the TU Dresden (project-id 247310070), and via the W\"urzburg-Dresden Cluster of Excellence on Complexity and Topology in Quantum Matter~--~\textit{ct.qmat} (EXC~2147, project-id 39085490). A.\,A. acknowledges financial support from the National Research Foundation (NRF) funded by the Ministry of Science of Korea (Grants: No.~2016K1A4A01922028, No.~2017R1D1A1B03033465, and No.~2019R1H1A2039733). S.\,E.\,N. acknowledges support from the International Max Planck Research School for Chemistry and Physics of Quantum Materials (IMPRS-CPQM). Research at Oak Ridge National Laboratory's Spallation Neutron Source was supported by the Scientific User Facilities Division, Office of Basic Energy Sciences, US Department of Energy.\vspace{-3pt}

\bibliographystyle{my-apsrev}
\bibliography{CeB6_FieldAngle}\vspace{-3pt}
\onecolumngrid

\clearpage
\renewcommand\thefigure{S\arabic{figure}}
\renewcommand\thetable{S\arabic{table}}
\renewcommand\theequation{S\arabic{equation}}
\renewcommand\bibsection{\section*{\sffamily\bfseries\footnotesize Supplementary References\vspace{-6pt}\hfill~}}

\citestyle{supplement}

\pagestyle{plain}
\makeatletter
\renewcommand{\@oddfoot}{\hfill\bf\scriptsize\textsf{S\thepage}}
\renewcommand{\@evenfoot}{\bf\scriptsize\textsf{S\thepage}\hfill}
\renewcommand{\fnum@figure}[1]{\figurename~\thefigure.~}

\makeatother
\setcounter{page}{1}\setcounter{figure}{0}\setcounter{table}{0}\setcounter{equation}{0}

\onecolumngrid\normalsize\vspace*{-3em}
\begin{center}
{\vspace*{0.1pt}\Large{Supplemental Material\smallskip\\\sl\textbf{``\hspace{1pt}Field-angle resolved magnetic excitations as a probe of\\hidden-order symmetry in CeB$_\text{6}$''}}}\vspace{-6pt}
\end{center}

\begin{figure*}[h]
\includegraphics[width=\textwidth]{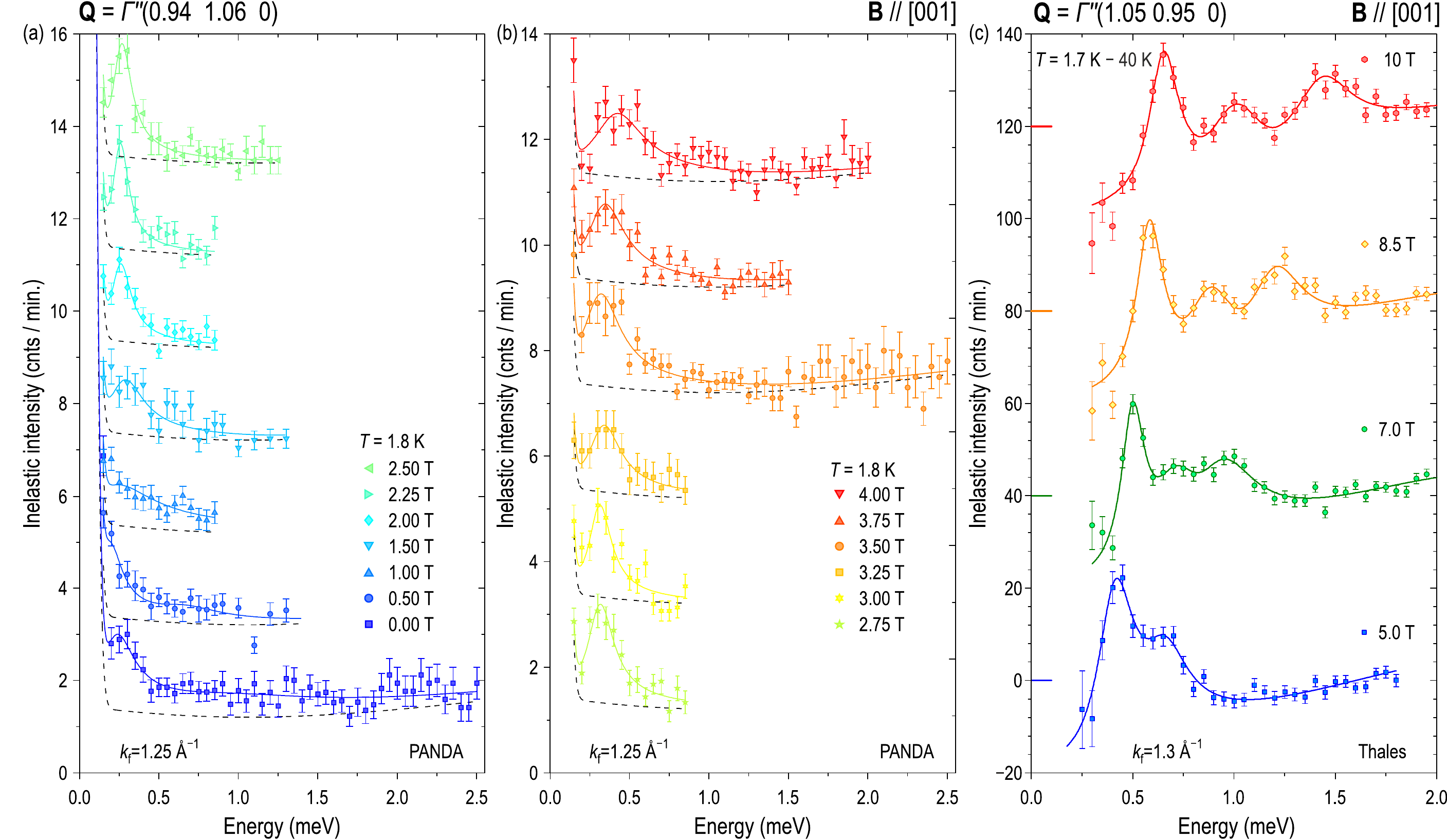}
\caption{INS spectra measured near the zone center $\Gamma''(110)$ for $\mathbf{B}\parallel[001]$. Panels (a,\,b) show raw data, whereas in panel (c), high-temperature ($T=40$~K) background has been subtracted.}
\label{Fig:GammaRawScans112and111}
\bigskip
\includegraphics[width=\textwidth]{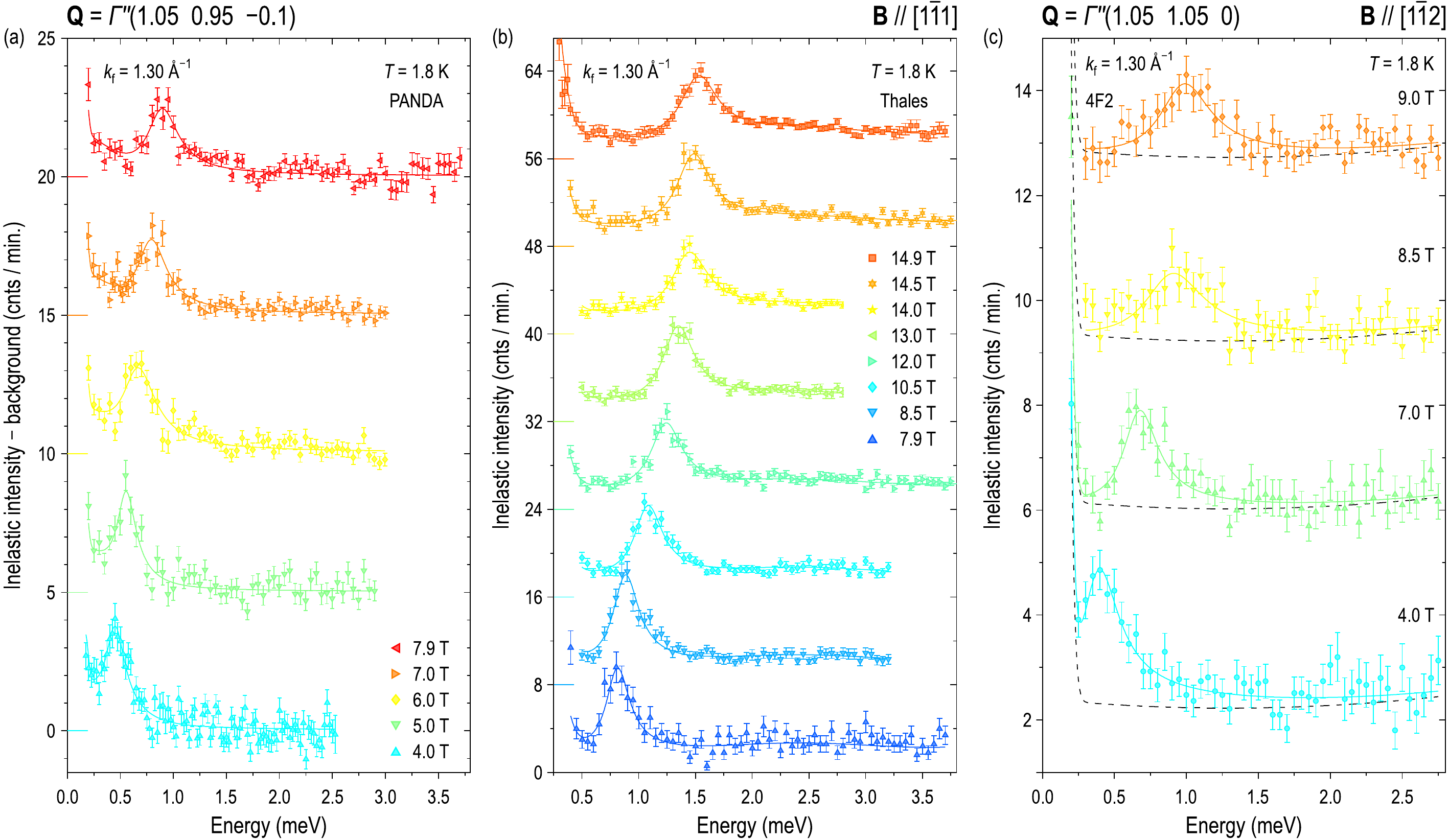}
\caption{INS spectra measured near the zone center $\Gamma''(110)$ for $\mathbf{B}\parallel[1\overline{1}1]$ and $[1\overline{1}2]$. In panel (a), the fitted background has been subtracted from the data.}
\label{Fig:GammaRawScans001}
\vspace{-3em}
\end{figure*}

\clearpage

\begin{figure}[t!]
\includegraphics[width=0.65\textwidth]{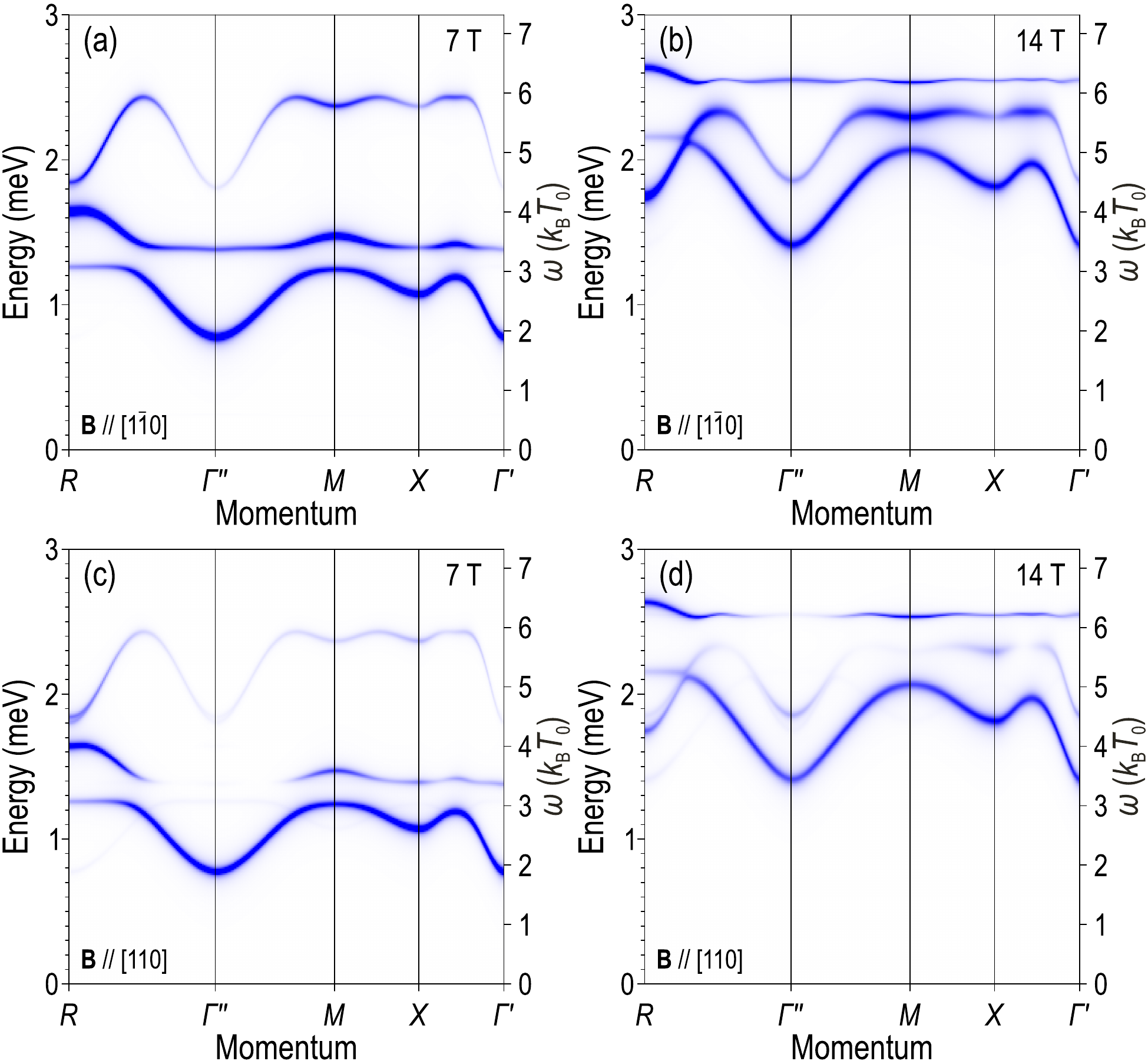}
\caption{The calculated dipolar structure function of magnetic excitations along the same polygonal path in momentum space as in Fig.~\ref{Fig:TheoryDisp} in the main text: $R(\frac{1}{2}\frac{1}{2}\frac{1}{2})$\,--\,$\Gamma''(110)$\,--\,$M(\frac{1}{2}\frac{1}{2}0)$\,--\,$X(\frac{1}{2}00)$\,--\,$\Gamma'(100)$. The top and bottom panels correspond to the two orthogonal magnetic field directions $[1\overline{1}0]$ and $[110]$, respectively. The calculations are shown for the magnetic fields $h'=1$ ($B\approx7$~T, left) and $h'=2$ ($B\approx14$~T, right).}
\label{Fig:TheoryB1m10vsB110}\bigskip
\includegraphics[width=0.65\textwidth]{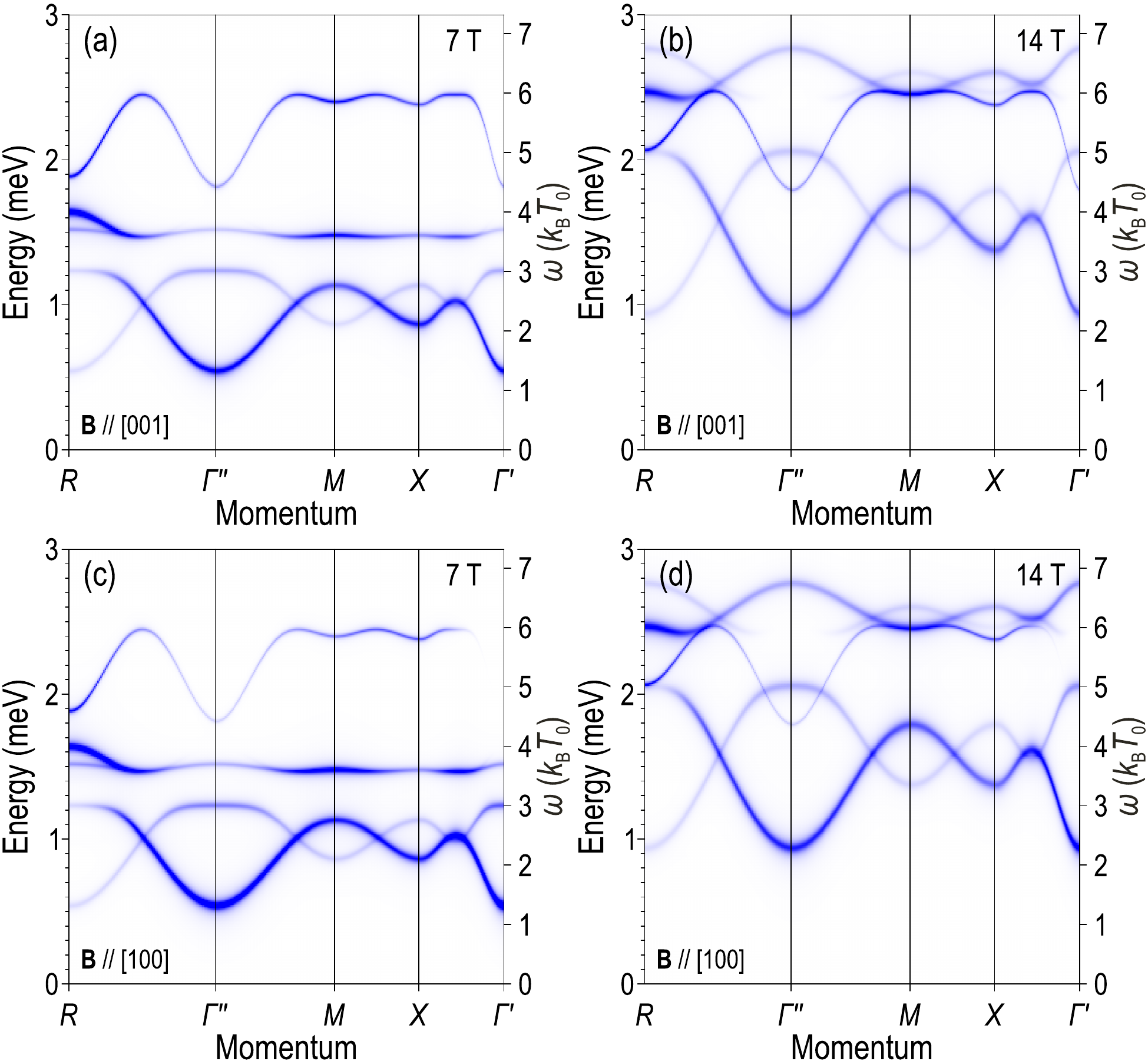}
\caption{The same as in the previous figure, but for magnetic field directions $[001]$ (top) and $[100]$ (bottom). Note the change in intensity of the upper mode near the $\Gamma'(100)$ point.}
\label{Fig:TheoryB100vsB001}
\end{figure}

\clearpage

\begin{figure}[t!]
\includegraphics[width=0.6\textwidth]{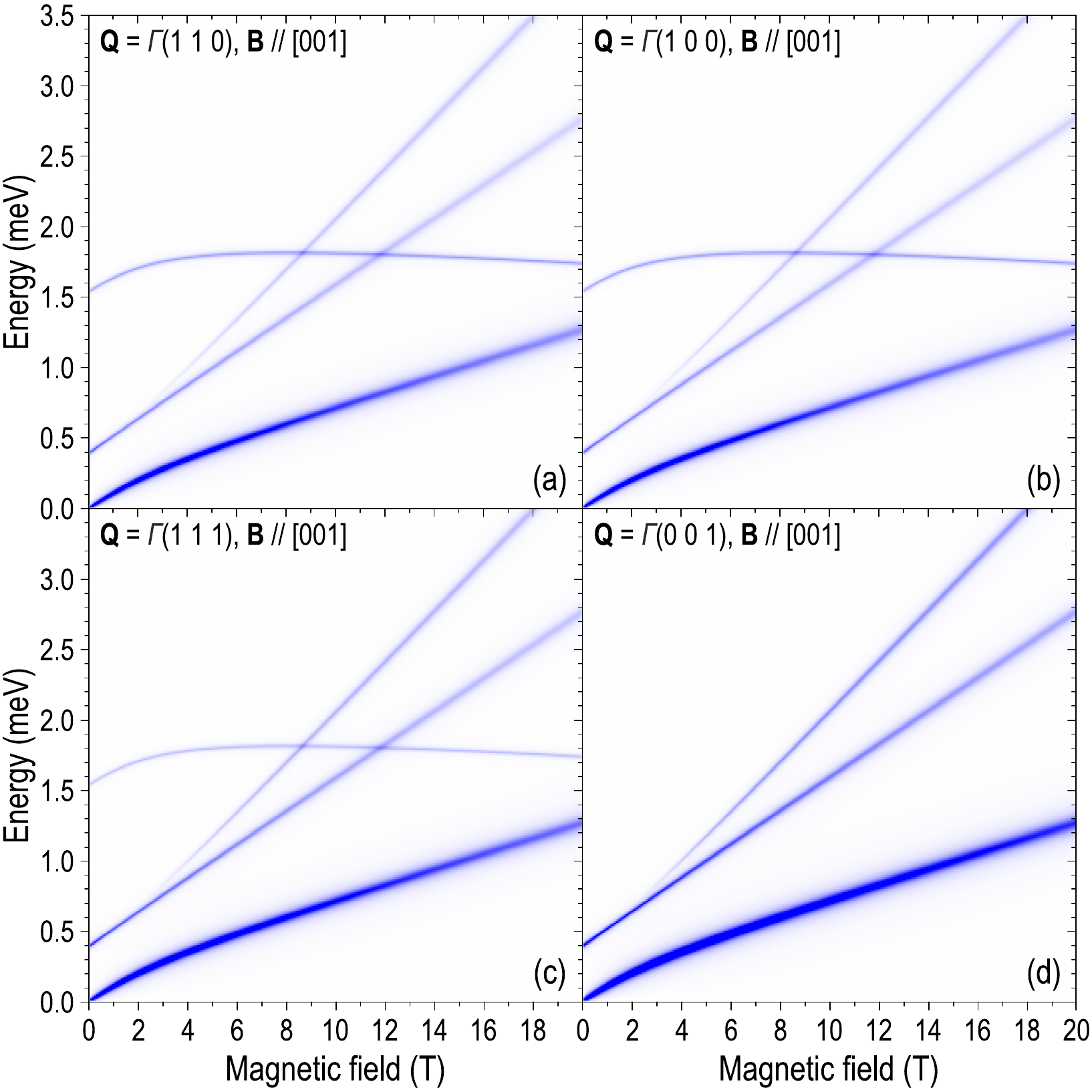}
\caption{Comparison of the field dependencies for the dynamic structure function at the $\Gamma$ point in a magnetic field applied along the $[001]$ cubic axis, calculated for different reciprocal-space points as indicated in the corresponding panels. In panels (a,\,b), the field is assumed to be perpendicular to the momentum-transfer vector, $\mathbf{B}\perp\mathbf{Q}$, in panel (d) the two vectors are parallel, $\mathbf{B}\parallel\mathbf{Q}$, whereas in panel~(c) the angle between the $\mathbf{Q}$ vector and the field direction is $\sim$\,46$^\circ$.}
\label{Fig:TheoryB001}\bigskip
\includegraphics[width=0.6\textwidth]{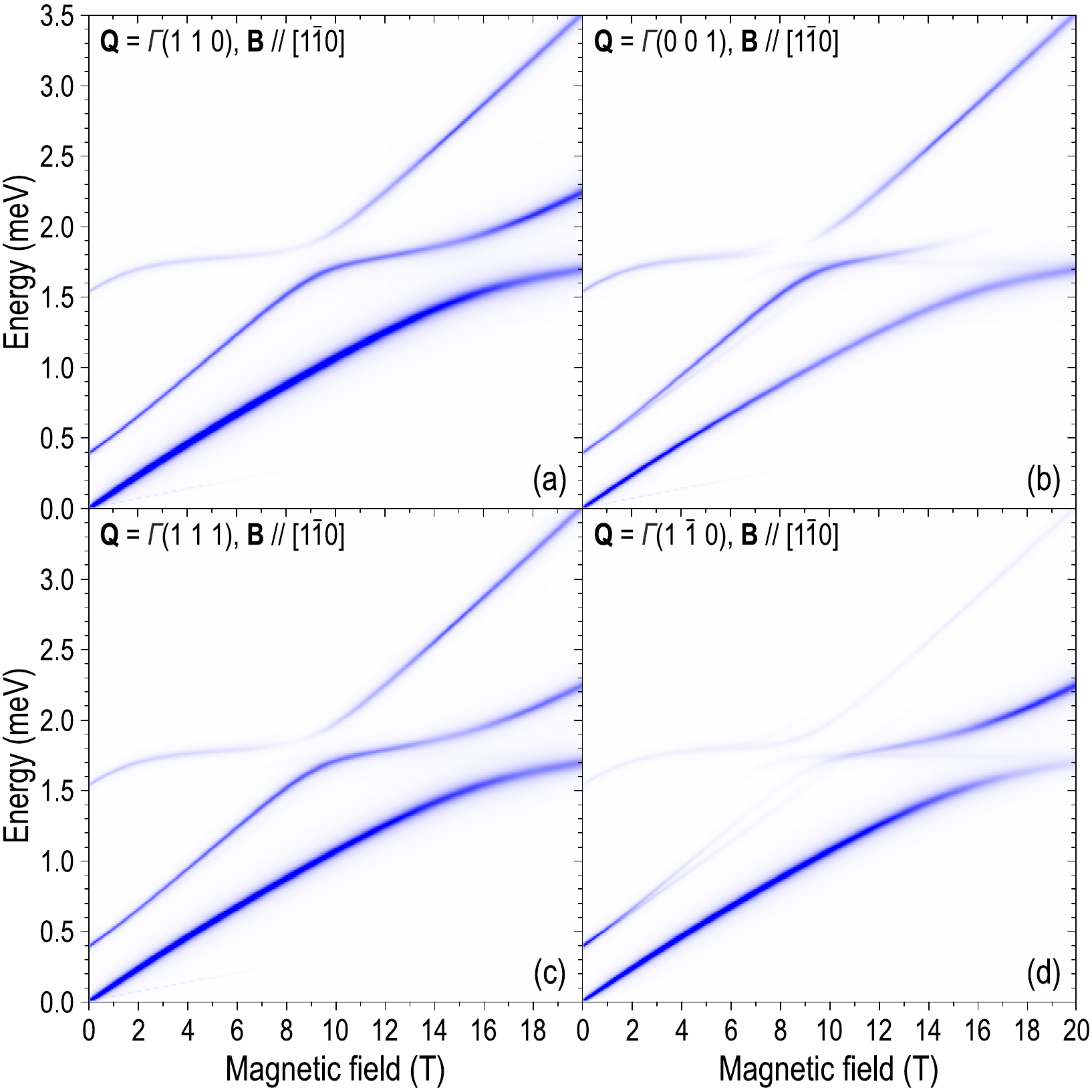}
\caption{Comparison of the field dependencies for the dynamic structure function at the $\Gamma$ point in a magnetic field applied along the $[1\overline{1}0]$ axis, calculated for different reciprocal-space points as indicated in the corresponding panels. In panels (a--c), the field is assumed to be perpendicular to the momentum-transfer vector, $\mathbf{B}\perp\mathbf{Q}$, whereas in panel~(d) the two vectors are parallel, $\mathbf{B}\parallel\mathbf{Q}$.}
\label{Fig:TheoryB110}
\end{figure}

\clearpage

\twocolumngrid\clearpage

\end{document}